\documentstyle[preprint,aps,epsfig]{revtex}

\begin{document}

\titlepage
\draft

\preprint{
\vbox{
\hbox{ADP-98-64/T331}
\hbox{IU/NTC 98-07}
}}

\title{Evidence for Charge Symmetry Violation in Parton Distributions} 

\normalsize
\author{C.Boros$^1$, J.T.Londergan$^2$ and A.W.Thomas$^1$}
\address {$^1$ Department of Physics and Mathematical Physics,\\ 
                and Special Research Center for the
                Subatomic Structure of Matter,\\ 
                University of Adelaide,
                Adelaide 5005, Australia}
\address{$^2$ Department of Physics and Nuclear
            Theory Center,\\ Indiana University,\\ 
            Bloomington, IN 47405, USA}
\date{\today}
\maketitle

\tighten 

\begin{abstract}  
By comparing structure functions measured in neutrino and  
charged lepton deep inelastic scattering, one can test the validity 
of parton charge symmetry. New experiments allow us to 
make such tests, which set rather tight upper limits on parton charge 
symmetry violation [CSV] for intermediate Bjorken $x$, but which appear 
to show sizable CSV effects at small $x$. 
We show that neither nuclear shadowing nor 
contributions from strange and antistrange quark distributions 
can account for the experimentally observed difference 
between the two structure functions.   
We are therefore forced to consider the possibility of a  
large CSV effect in the nucleon sea quark
distributions. We discuss the consequences of this effect for  
other observables, and we propose an experiment which could detect 
a large CSV component in the nucleon sea.

\end{abstract}


\newpage 

\section{Introduction} 

In discussing the strong interaction, it is customary to assume 
the validity of charge symmetry, which interchanges protons and 
neutrons (simultaneously interchanging up and down quarks).  For example, 
all phenomenological analyses of deep inelastic scattering 
data in terms of parton distribution functions assume  
charge symmetry from the beginning.  Our faith in charge 
symmetry is justified from our experience in 
nuclear physics, where this symmetry 
is respected to a high degree of precision.  Most experimental 
low-energy tests of charge symmetry find that it is good to at least
$1\%$ in reaction amplitudes \cite{Miller,Henley}. 
Until recently such an assumption seemed to be 
justified, as there was no compelling experimental evidence against 
parton charge symmetry.  The quantitative evidence which could be 
extracted from high energy experiments, although not particularly precise, 
was consistent with charge symmetric parton distributions 
\cite{Lon98}.  

Experimental verification of charge symmetry is difficult, partly 
because the relative charge symmetry violation (CSV) effects 
are expected to be small, requiring high precision experiments to 
measure CSV effects, 
and partly because CSV often mixes with parton flavor symmetry violation 
(FSV).  Recent experimental measurements by the NMC Collaboration 
\cite{NMCfsv}, demonstrating the violation of the Gottfried 
sum rule \cite{Gottfried}, have been widely interpreted as 
evidence for what is termed SU(2) FSV. The measurement of the  
ratio of Drell-Yan cross sections in proton-deuteron and 
proton-proton scattering, first by the NA51-Collaboration at CERN 
\cite{Na51} and more recently by the E866 experiment 
at FNAL \cite{E866}, also indicate substantial FSV. 
However,  both of these experiments 
could in principle be explained by sufficiently large CSV effects 
\cite{Ma1,Steffens}, even in the limit of exact flavor symmetry.  
In view of these ambiguities in the interpretation of current 
experimental data, it would be highly desirable to have experiments 
which separate CSV from FSV.  
A few experiments have been already proposed 
\cite{Tim1,Tim2} and could be carried out in the near future. 

Recent experiments now allow us for the first time to make
precision tests which could put tight upper limits on parton 
CSV contributions.  The NMC measurements of muon DIS on deuterium 
\cite{NMC} provide values for the charged lepton structure
function $F_2^\mu(x,Q^2)$.  In a similar $Q^2$ regime the 
CCFR Collaboration \cite{CCFR} extract the structure functions 
$F_2^\nu(x,Q^2)$ from neutrino--induced charge-changing reactions.  
As we show in sec.\ II, the ``charge ratio'', 
which can be constructed from these two quantities (plus 
information about the strange quark distribution) can in 
principle place strong constraints on parton CSV distributions.  
We will show that, for intermediate values $ 0.1 \le x \le 0.4$, 
the agreement between the two structure functions is impressive, 
and provides the best upper limit to date on parton CSV terms.  
However, the charge ratio shows a substantial 
deviation from unity in the region $x < 0.1$, which might suggest 
surprisingly 
large charge symmetry violation.  In a recent Letter \cite{Bor98} we 
argued that the data supported this conclusion.  However,   
several important corrections have to be applied to the data 
before any conclusions can be reached.  These corrections are
especially important for the neutrino cross sections.    

In sec.\ II we discuss the uncertainties involved in the 
analysis of the data.  Most corrections have already been accounted
for in the present experimental analysis.  We particularly focus on 
two aspects of the neutrino reactions: heavy target 
corrections and effects due to strange and antistrange quark distributions. 
In sec.\ III we demonstrate that neither of these effects are 
sufficient to account for the apparent discrepancy at small $x$.  
The charge symmetry violating distributions can be obtained from a 
combination of neutrino charged current structure functions, muon
structure functions and strange quark distributions extracted from 
dimuon production in neutrino reactions.  We construct such a 
combination and extract the CSV terms.  Assuming the validity of 
the experimental data, we find CSV effects on the order of 25\% of 
the sea quark distributions at low $x$. In sec.\ IV 
we discuss the consequences of 
such large CSV effects on other observables.  We   
examine  the role played by CSV in the extraction of the FSV 
ratio $\bar d/\bar u$, in the Gottfried sum rule 
and in the experimental determination of the Weinberg 
angle sin$^2\theta_W$.  
In sec.\ V we suggest an experiment which could measure the substantial 
CSV suggested by our analysis. 

\section{Comparing Structure Functions From Neutrino and Charged 
Lepton Reactions}

Our analysis of parton charge symmetry violation is based on
the ``charge ratio,'' which we review here.  This depends on
the ratio of $F_2$ structure functions extracted from charged
lepton reactions with those from neutrino charge--changing
reactions.  Because neutrino cross sections are so small, at 
present the structure functions can only be measured for 
heavy targets such as iron.  Furthermore, in order to obtain
useful statistics, the data must be integrated over all 
energies for a given $x$ and $Q^2$ bin.  As a result, only 
certain linear combinations of neutrino and antineutrino 
structure functions can be obtained.  The process by which 
we attempt to extract parton CSV contributions is complicated, 
and requires input from several experiments.  In this section
we review this process in detail. 

\subsection{The ``Charge Ratio'' and Charge Symmetry Violation}

Structure functions measured in neutrino and muon deep inelastic 
scattering are interpreted in terms of parton distribution functions. 
Since the operation of charge symmetry maps up quarks to down quarks, 
and protons to neutrons, at the level of parton distributions, 
charge symmetry implies the equivalence between up (down) 
quark distributions in the proton and down (up) quark distributions 
in the neutron.  In order to take CSV in 
the parton distributions into account, we define 
the charge symmetry violating distributions as 
\begin{eqnarray} 
\delta u(x)& =&  u^p(x) -d^n(x) \nonumber\\ 
\delta d(x)& =&  d^p(x) -u^n(x), 
\end{eqnarray} 
where the superscripts $p$ and $n$ refer to 
quark distributions in the proton and neutron, respectively. 
The relations for CSV in antiquark distributions 
are analogous.  If charge symmetry were exact then the 
quantities $\delta u(x)$ and $\delta d(x)$ would vanish. 

In the quark-parton model the structure functions measured in 
neutrino, antineutrino  and charged lepton DIS  
on an isoscalar target, $N_0$,    
are given in terms of the  
parton distribution functions and the 
charge symmetry violating distributions defined above by 
\cite{Lon98}   
\begin{eqnarray}
  F_2^{\nu N_0} (x,Q^2) &=& x[u(x)+ \bar u(x) +d(x) +\bar d(x)
     + 2 s(x) + 2 \bar c(x) -\delta u(x)-\delta \bar d(x)] \nonumber \\ 
  F_2^{\bar \nu N_0} (x,Q^2) &=& x[u(x)+ \bar u(x) +d(x) +\bar d(x)
     + 2 \bar s(x) + 2  c(x) -\delta d(x)-\delta \bar u(x)] \nonumber \\
  xF_3^{\nu N_0}(x,Q^2) &=& x[u(x) + d(x) -\bar u(x) - \bar d(x) 
       +2 s(x)-2 \bar c(x) -\delta u(x) +\delta\bar d(x)]   
      \nonumber\\ 
  xF_3^{\bar\nu N_0}(x,Q^2) &=& x[u(x) + d(x) -\bar u(x) - \bar d(x) 
       -2 \bar s(x)+2  c(x) -\delta d(x) +\delta\bar u(x)]  \nonumber\\ 
  F_2^{\ell N_0}(x,Q^2) & =& \frac{5}{18} x 
   [ u(x) + \bar u(x)
 +d(x) +\bar d(x) + \frac{2}{5} (s(x) + \bar s(x)) 
  +  \frac{8}{5}(c(x)+\bar c(x)) \nonumber\\ 
 & -  & \frac{4}{5} 
 (\delta d(x)+\delta \bar d(x)) - \frac{1}{5} ( \delta u(x)+\delta 
   \bar u(x))]
\label{eq2} 
\end{eqnarray}
Here, and in the following, quark distributions without 
superscripts denote quark distributions in the proton.  From 
now on, we will disregard charm quark contributions to the
structure functions.  
Since phenomenological parton distribution functions assume 
the validity of charge symmetry, possible CSV effects are folded 
into the commonly used phenomenological parton 
distribution functions in a highly 
non-trivial way. Nevertheless,  using the above relations, it is 
possible to test the validity of charge symmetry  by 
building appropriate linear combinations or ratios of the measured  
structure functions. One such  possibility is to 
calculate the ``charge ratio'',  which relates the neutrino 
structure function to the structure function measured in charged 
lepton deep-inelastic scattering
\begin{eqnarray}
 R_c(x,Q^2) & \equiv  & \frac{F_2^{\mu N_0}(x,Q^2)}{\frac{5}{18}
 F_2^{\nu N_0}(x,Q^2) -x( s(x) +\bar s(x))/6} \nonumber\\ 
&\approx & 1 - \frac{s(x) -\bar s(x)}{\overline{Q}(x)} + 
 \frac{4\delta u(x) - \delta \bar u(x) - 4 \delta d(x) 
+\delta \bar d(x)}{5 \overline{Q}(x)}.   
\label{rc}
\end{eqnarray}
Here, we defined 
$\overline{Q}(x) \equiv \sum_{q=u,d,s} (q(x)+\bar q(x))-
3(s(x)+\bar s(x))/5$, 
and we have expanded Eq.\ \ref{rc} to lowest order in small quantities.  
From Eq.\ \ref{rc} we see that any deviation of the charge ratio from 
unity, at any value of $x$, would be due either to CSV effects or to 
different strange and antistrange quark distributions.  Analogous
relations could be obtained using structure functions from antineutrinos, 
or from a linear combination of neutrino and antineutrino structure
functions.  For example, we can derive  
\begin{eqnarray}
 {\cal R}_c (x,Q^2) & \equiv  & 
 \frac{F_2^{\mu N_0}(x,Q^2)}{\frac{5}{18}
 {\cal F}_2^{\nu N_0}(x,Q^2) -x( s(x) +\bar s(x))/6} \nonumber\\ 
&\approx & 1 + \frac{3\left( \delta u(x) + \delta \bar u(x) - \delta d(x) 
-\delta \bar d(x)\right)}{10 \overline{Q}(x)}.   
\label{rc2}
\end{eqnarray}
In Eq.\ \ref{rc2} ${\cal F}_2^{\nu N_0}(x,Q^2) = (F_2^{\nu N_0}(x,Q^2) + 
 F_2^{\bar\nu N_0}(x,Q^2))/2$ is the average of the structure functions
from neutrino and antineutrino reactions; deviations from one in 
the ratio ${\cal R}_c (x)$ depend only on parton CSV 
contributions, and have no contribution from strange or antistrange 
quark distributions. 

\begin{figure}
\epsfig{figure=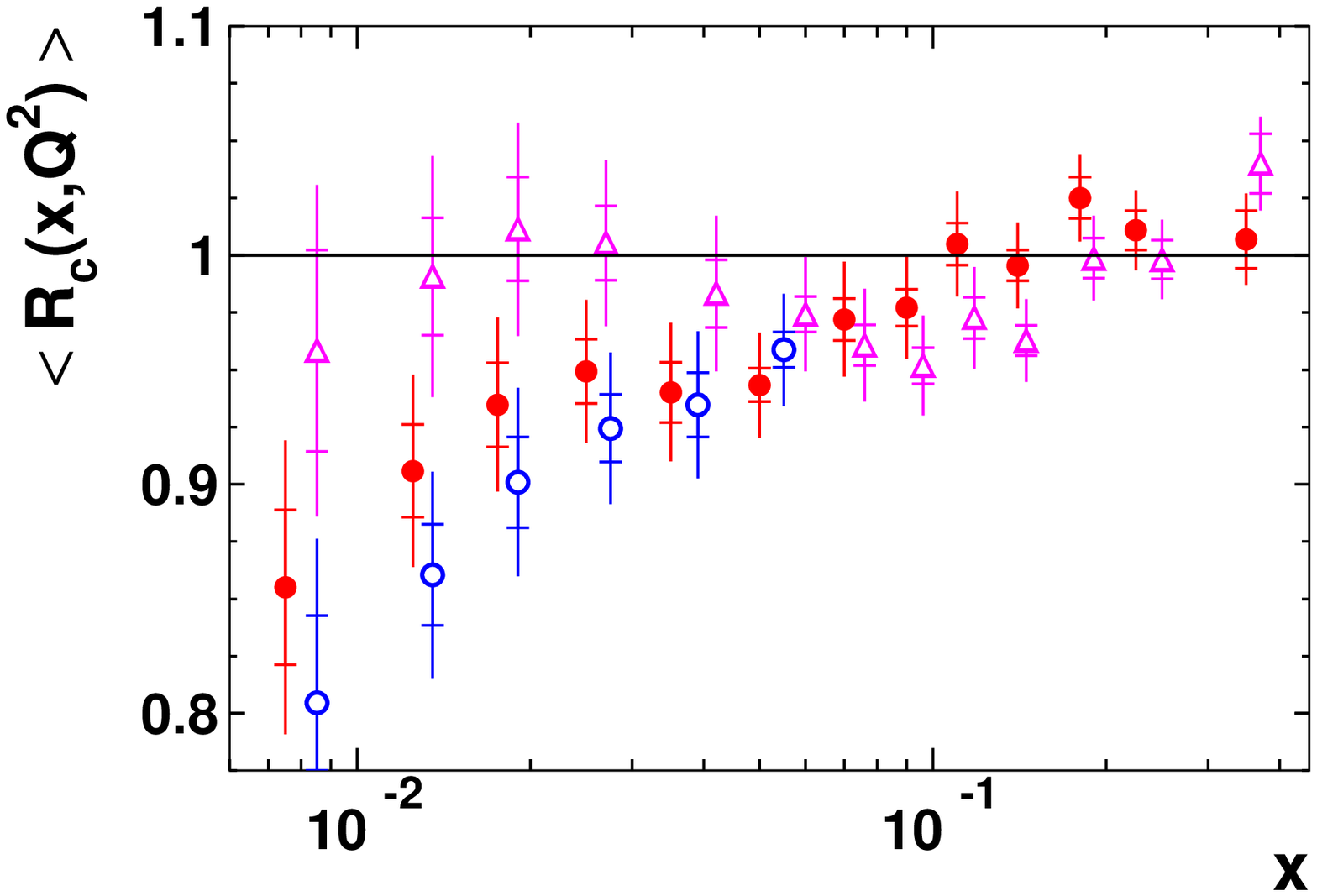,height=12.cm}
\caption{The ``charge ratio'' $R_c$ of Eq.\ \protect\ref{rc} vs.\ $x$ 
	calculated using CCFR \protect\cite{CCFR} data for neutrino 
        and NMC \protect\cite{NMC} data for muon
         structure functions. Open triangles: no heavy target
         corrections; open circles: $\nu$ data corrected for heavy
         target effects using corrections from charged lepton scattering;
         solid circles: $\nu$ shadowing corrections calculated in the
         ``two phase'' model.
          Both statistical and systematic errors are shown. }
\label{fig1}
\end{figure}

The recent measurement of the structure function
$F_2^\nu$ by the CCFR-Collaboration \cite{CCFR} makes it possible
to carry out a precise comparison between $F^\nu_2(x,Q^2)$ 
and $F_2^{\mu}(x,Q^2)$ for the first time.    
The CCFR-Collaboration compared the neutrino structure function
$F_2^\nu (x,Q^2)$ extracted from their data on an iron target \cite{CCFR}
with $F_2^\mu (x,Q^2)$ measured for the deuteron by the NMC
Collaboration \cite{NMC}.  In the region of intermediate values of
Bjorken $x$ ($0.1 \le x \le 0.4$), they found very good agreement between
the two structure functions.  In this $x$ region, this allows us to set 
upper limits of a few percent on parton CSV contributions.  
On the other hand, in the small
$x$-region ($x < 0.1$), the CCFR group found that the two
structure functions differ by as much as 10-15$\%$. 
This can be seen in Fig.\ref{fig1} where the ``charge ratio'' 
has been obtained by integrating over the region of overlap in  
$Q^2$ of the two experiments.  The open and solid circles in 
Fig.\ \ref{fig1} represent two different ways of calculating 
nuclear shadowing corrections, as we will discuss later.  
 
\subsection{Extracting Structure Functions From Neutrino Cross
Sections}

In order to perform tests of parton distributions through, say, 
the charge ratio of Eq.\ \ref{rc}, we need the structure functions
from neutrino charge--changing reactions on a free proton
and neutron.  These are written in terms of parton distributions
in Eq.\ \ref{eq2}.  Because of the extremely small cross sections
for neutrino--induced reactions, we are able to 
obtain statistically meaningful cross sections only from heavier 
targets such as iron.  We then have to make the following 
corrections in order to extract the neutrino structure functions on 
``free'' nucleons, averaged over proton and neutron, 
$F_j^{\nu,\bar\nu\,N_0}(x,Q^2)$, ($j=2,3$): i) The nuclear structure 
functions $F_j^{\nu,\bar\nu\,Fe}(x,Q^2)$ must be extracted from the 
cross sections; ii) The nuclear structure functions need to be corrected 
for the excess of neutrons in iron (isoscalar effects); 
iii) Kinematic corrections must be applied to account for
heavy quark thresholds, particularly charm quark threshold effects  
(Eq.\ \ref{rc} is valid only well above all heavy quark thresholds); 
iv) Heavy target corrections must be applied, to  
convert structure functions for nuclei to those for free protons
and neutrons; v) The neutrino and muon cross sections must be 
properly normalized.  In order to test charge symmetry, all these 
corrections have to be taken 
into account. The data have already been 
corrected for normalization, isoscalar and charm threshold effects by the 
CCFR-Collaboration in their analyses \cite{CCFR}.  There is a
thorough discussion of these points in the thesis by W. Seligman 
\cite{Sel97}.  Here we will review how the nuclear structure functions
are extracted from the cross sections, the heavy target corrections
for neutrino reactions, and the role of both strange quarks and 
CSV effects in neutrino structure functions.  

The cross sections for neutrino and antineutrino scattering on a 
nuclear target containing $A$ nucleons can be written as   
\begin{equation}
\frac{d\sigma^{\nu ,\bar\nu\,A}}{dxdQ^2} = 
  \frac{G_F^2}{2\pi x} [ \frac{1}{2}(F_2^{\nu ,\bar\nu\,A}(x,Q^2)
   \pm xF_3^{\nu ,\bar\nu\,A} (x,Q^2))
     + \frac{\xi^2}{2}(F_2^{\nu ,\bar\nu\,A} (x,Q^2)\mp
x F_3^{\nu ,\bar\nu\,A} (x,Q^2)) ]. 
\label{nuxsec}
\end{equation} 
In Eq.\ \ref{nuxsec} the upper (lower) sign is associated with neutrino
(antineutrino) cross sections.  We have assumed the validity of the 
Callan-Gross relation and neglected terms of order $Mxy/2E$, and
we introduced the variable $\xi=(1-y)$.  It would be straightforward
to remove these assumptions.  With a large enough count rate, the 
$x$ and $y$ dependence of the cross sections could be separately 
measured.  By plotting the measured differential 
cross sections for fixed $x$ and $Q^2$ as a function of $\xi^2$, the 
structure functions $F_2$ and $F_3$ can be determined from the slopes  
and intercepts of the resulting straight lines. 
The crucial question is, of course, whether the statistics 
of the experiment are sufficient for the structure functions to 
be extracted in this way. 

\begin{figure}
\epsfig{figure=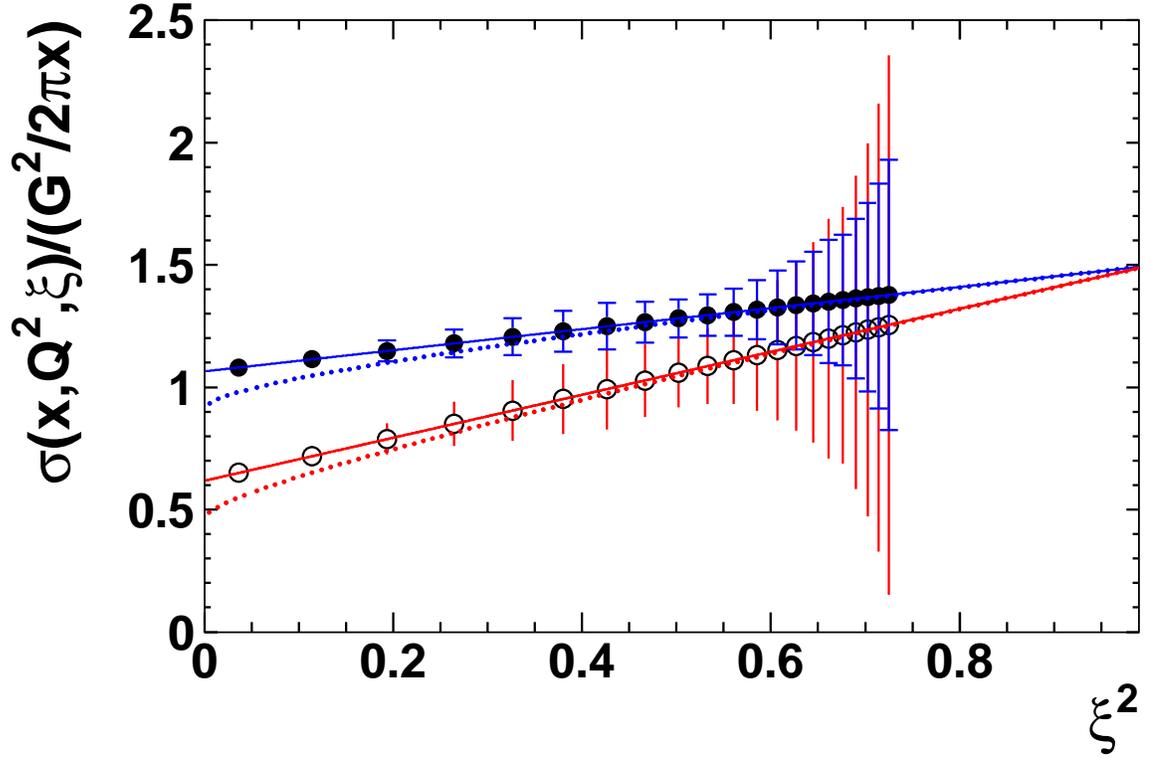,height=12.cm}
\caption{The differential cross sections for neutrino  
   (solid circles) and antineutrino (open circles)  
   deep inelastic scattering as a function of the variable 
  $\xi^2\equiv (1-y)^2$ for $x=0.03$ and $Q^2=4$ GeV$^2$.   
  The solid and dotted lines are the results with and without the 
  Callan-Gross relation, respectively. The statistical errors are 
  estimated using the experimental fluxes of neutrinos  
  and antineutrinos.}
\label{fig2}
\end{figure}

To illustrate this problem we calculated the statistical 
errors in each energy bin. For this calculation, 
we used the experimental determined 
fluxes, the total and differential neutrino and antineutrino cross sections 
to obtain the expected number of events in a given $x$, $Q^2$ 
and energy bin. We estimated the statistical errors using 
$\Delta\sigma =\sigma/\sqrt{N}$. In Fig.\ref{fig2}   
$\sigma^{\nu ,\bar\nu} (x,Q^2,\xi^2)/(G_F^2/2\pi x)$  
is plotted as a function of $\xi^2$.  
The solid lines are the results using the CTEQ parton
distribution functions and assuming the validity of the 
Callan-Gross relation. The dotted lines are the results obtained 
without using the Callan-Gross relation. Here, we used the
parametrization of Whitlow \cite{Whi90b} for the ratio of the longitudinal
and transverse photo-absorption cross sections.  The current
statistics do not allow one to extract the individual structure
functions.  The error bars represent the expected statistical errors.  
An order of magnitude more events would be necessary to decrease the 
statistical errors sufficiently that one could consider extracting 
the structure functions directly, and systematic errors would
further complicate this analysis.   

Since the number of events is so small that individual structure 
functions cannot be extracted from the data, the cross sections
in a given $x$ and $Q^2$ bin are integrated over all energies.  
After this integration is performed, Eq.\ \ref{nuxsec} can be
written as two linear equations, one for neutrino and the other
for antineutrino events: 
\begin{eqnarray} 
N^\nu(x,Q^2)  &=& A_2^\nu \, \,F_2^{\nu Fe}(x,Q^2) + A_3^\nu \, 
xF_3^{\nu Fe}(x,Q^2) \nonumber \\ 
  N^{\bar\nu}(x,Q^2) &=& A_2^{\bar\nu}\,F_2^{\bar\nu Fe}(x,Q^2) - 
  A_3^{\bar\nu}\,xF_3^{\bar\nu Fe}(x,Q^2) ~~~. 
\label{nneut}
\end{eqnarray}
In Eq.\ \ref{nneut} $N^\nu$ ($N^{\bar\nu}$) is the number of 
neutrino (antineutrino) events in a given $x$ and $Q^2$-bin integrated 
over the incident neutrino and antineutrino energies. 
$A^\nu_i$ and $A_i^{\bar\nu}$ ($i=2,3$)  
represent the coefficients, $A_i(y)$, of the structure functions 
multiplied by the neutrino and antineutrino fluxes,  
$\Phi^\nu (E)$ and $\Phi^{\bar\nu} (E)$, respectively, and integrated
over all energies        
\begin{eqnarray} 
A^\nu_i &=& \int dE\, A_i(y)\, \Phi^\nu (E) \nonumber \\ 
A^{\bar\nu}_i & = &\int  dE\, A_i(y)\, \Phi^{\bar\nu} (E) . 
\end{eqnarray}   

The individual structure functions for neutrino and antineutrino 
reactions are extracted by taking linear combinations of the
relations in Eq.\ \ref{nneut} and making corrections using 
phenomenological parton distribution functions.  For example, 
from Eq.\ \ref{eq2} we see that for an isoscalar target, 
$F_2^{\nu\,N_0}(x,Q^2)=F_2^{\bar\nu\,N_0}(x,Q^2)$ 
if charge
symmetry is valid and $s(x) = \bar{s}(x)$.  Thus we can form
linear combinations of the terms in Eq.\ \ref{nneut} such that
these terms cancel and we are left only with the $F_3$ structure
functions.  Similarly, assuming charge symmetry we have 
\[ F_3^{\nu\,N_0}(x,Q^2) - F_3^{\bar\nu\,N_0}(x,Q^2) = 2[s(x) - 
\bar{s}(x)]  ~~. \] 
We can then take a linear combination of the terms in Eq.\ \ref{nneut} 
which gives this function.  If the strange quark distribution is
taken from a phenomenological model, we can extract a linear
combination of the $F_2$ structure functions for neutrinos and
antineutrinos on a nuclear target.  

We will discuss how the structure functions are extracted, and 
particularly the role of CSV and strange quark distributions in 
this process.  However, at this stage we review how heavy target
corrections are calculated, in order to extract the structure
functions for free nucleons from those measured on a heavy
nuclear target.  

\subsection{Heavy Target Corrections in Neutrino Reactions}

As is well known, the structure functions measured on heavy
targets are not equal to those observed for light targets such 
as the deuteron.  At small $x$ values, nuclear shadowing effects 
play a major role; at large $x$, nuclear Fermi motion effects 
dominate, and at intermediate $x$ ``EMC'' effects play a 
significant role \cite{Arneodo}.   Such effects have been 
systematically measured in charged lepton reactions.  

In analyzing neutrino scattering data, it is generally assumed 
that heavy target corrections will be the same as those observed 
in charged lepton reactions.  {\it A priori}, there is no reason to assume   
that neutrino and charged lepton heavy target corrections should be 
identical.  Heavy target corrections for neutrinos are generally 
applied by multiplying the experimental structure functions at 
a given $x$ value by
the quantity $R\equiv F^{\ell A}_{2}(x,Q^2)/F^{\ell D}_{2}(x,Q^2)$, 
the ratio between the $F_2$ structure function measured on heavy targets
and that of the deuteron for charged lepton deep inelastic scattering, 
at the same $x$ value.
However, as is well known, shadowing corrections are very much
$Q^2$ dependent for smaller $Q^2$ values (where a considerable part
of the available data was taken), and the $Q^2$ and $x$-dependence of
the data are strongly correlated because of the
fixed target nature of these experiments.

We re-examined heavy target corrections to deep-inelastic neutrino 
scattering, focusing on the differences between neutrino and charged  
lepton scattering and on effects due to the $Q^2$-dependence of 
shadowing for moderately large $Q^2$.  This work will be
published elsewhere \cite{Boros}; here we briefly review the results
of that work.  We used a two phase model  
which has been successfully applied to the description of shadowing 
in charged lepton DIS  
\cite{Badelek,Melni}. In this approach,  
vector meson dominance is used to describe the low $Q^2$ virtual 
photon or $W$ interactions, and Pomeron exchange is used for the 
approximate scaling region. In generalizing this approach 
to weak currents, the essential differences in shadowing 
between neutrino and charged lepton 
deep inelastic scattering are:  
(i) the axial-vector current is only partially conserved, 
in contrast to the vector current; and (ii) the weak current 
couples not only to vector but also to axial vector mesons 
\cite{Stodolsky,VMD,Bell,Boris}. 

Partial conservation of the axial current (PCAC) requires   
that the divergence of the axial current does not vanish 
but is proportional to the pion field for $Q^2=0$. This 
is Adler's theorem \cite{Adler}, 
which relates the neutrino cross section to the pion 
cross section  on the same target for $Q^2=0$. 
Thus, for low $Q^2$($\approx m_\pi^2$) shadowing in neutrino 
scattering is determined by the absorption of  pions on the target. 
For larger $Q^2$-values the contributions of vector and axial vector mesons 
become important. The coupling of the weak current 
to the vector and axial 
vector mesons and that of the electro-magnetic current 
to vector mesons are related to each other by the ``Weinberg sum rule'' 
$f_{\rho^+}^2=f_{a_1}^2=2f_{\rho^0}^2$. 
Since the coupling of the vector (axial vector) 
mesons to the weak current is  twice as large as  the coupling 
to the electro-magnetic current, but the structure function is 
larger by a factor of $\sim18/5$  in the neutrino 
case, we expect that shadowing due to VMD in neutrino reactions is 
roughly half of that in charged lepton scattering.  

For larger $Q^2$-values, shadowing due to Pomeron exchange between the 
projectile and two or more constituent nucleons dominates. Since 
Pomeron-exchange models the interaction between partons 
in different nucleons and the scattering of the $W$ takes 
place on only one parton, this processes is of leading twist in 
contrast to the VMD and pion contributions. 
The coupling is given by the coupling 
of the photon or $W$ to the quarks in the exchanged Pomeron. 
It changes in the same way as the structure function does 
in switching from neutrino to charged lepton 
scattering. Thus, for large $Q^2$ values ($>10$ GeV$^2$), shadowing in both
cases should have approximately the same magnitude. 
In the intermediate $Q^2$-region ($1<Q^2<10$ GeV$^2$), where VMD 
is relatively important, we expect to see differences between  shadowing 
in neutrino and charged lepton scattering.  
We recall that this is precisely the region where 
the discrepancy between CCFR and NMC is significant.  
There are also nuclear effects in the deuteron. However,
because of the low density of the deuteron,
these are (relatively speaking)
very small and have a negligible effect on the charge ratio.

We calculated the shadowing corrections to the CCFR neutrino 
data using the two-phase model of Ref.\cite{Badelek,Melni}.  With 
this corrected CCFR data, we calculated the charge ratio $R_c$ of 
Eq.\ \ref{rc} between CCFR and NMC data. The result is shown in 
Fig.\ref{fig1}.  The open triangles show the charge ratio when no
shadowing corrections are used.  The open 
circles show the charge ratio when heavy target shadowing corrections
from charged lepton reactions are applied to the neutrino data, 
and the solid circles show the result when the neutrino shadowing 
corrections from our two-phase model are applied.  
At small $x$, using the ``correct'' neutrino shadowing 
corrections reduces the deviation of the charge ratio from unity.   
Nevertheless, the charge
ratio is still not compatible with one at small $x$.  
In summary, properly accounting for shadowing corrections in the 
neutrino structure function decreases, but does not resolve, 
the low-$x$ discrepancy between the CCFR and the NMC data. 

\subsection{Strange Quark and CSV Contributions to Structure Functions}

In Eq.\ \ref{nneut} we showed that, after integrating neutrino
charged--current cross sections over all energies, we obtain two
equations in four unknowns, the structure functions $F_2$ and 
$F_3$ for neutrino and antineutrino reactions.  If the neutrino and 
antineutrino structure functions were equal, $F_2^{\nu Fe}(x,Q^2)=
F_2^{\bar\nu Fe}(x,Q^2)$, 
with an analogous relation for $xF_3^{\nu Fe}(x,Q^2)$, 
then Eq.\ \ref{nneut} would provide two linear equations in two unknowns. 
As we discussed previously, several corrections need 
to be applied before we can extract the structure functions on 
a ``free'' isoscalar target $N_0$, and compare the structure functions 
to the parton distributions given in Eq.\ \ref{eq2}.  First, 
since iron is not an isoscalar target we need to make corrections for 
the excess neutrons. Next, we need to estimate the contributions 
from strange quark distributions and charge symmetry violating 
parton distributions.  Finally, we need to make heavy target corrections
as reviewed in the preceding section.  

We begin by splitting the neutrino and antineutrino structure 
functions on iron into isoscalar and non-isoscalar 
parts.  For a target with $Z$ protons and $N=A-Z$ neutrons we define 
the quantity $\beta\equiv (N-Z)/A$: 
\begin{equation} 
  F_i^{\nu ,\bar\nu Fe}= 
\frac{1}{2} [F_i^{\nu ,\bar\nu p} + F_i^{\nu ,\bar\nu n}] 
-\frac{\beta}{2} [F_i^{\nu ,\bar\nu p} - F_i^{\nu ,\bar\nu n}]\, .  
\label{s12}   
\end{equation} 
The first term on the right of Eq. \ref{s12}  
corresponds to the neutrino and antineutrino structure functions 
on an isoscalar target, $N_0$. 
The second terms include corrections arising from the 
non-isoscalarity of the target. 
In the absence of CSV, these corrections are basically given by 
the difference  between  up and down valence quark distributions  
and  have been  taken into account in the extraction 
of the structure functions. However, the non-isoscalarity of the target 
leads also to CSV corrections. 

We define the sum and difference of the neutrino and 
antineutrino structure functions on a target $A$ as 
\begin{eqnarray} 
  {\cal F}_i^A & \equiv& \frac{1}{2} [ F_i^{\nu A} + F_i^{\bar\nu A}]
  ~~, \nonumber \\ 
  \Delta {\cal F}_i^A & \equiv& \frac{1}{2} [ F_i^{\nu A} - 
  F_i^{\bar\nu A}] ~~; 
\end{eqnarray}
the structure functions $F_i^{\nu ,\bar\nu Fe}$ can then be written as 
\begin{equation}
  F_i^{\nu ,\bar\nu Fe} =  
      {\cal F}_i^{N_0} \pm \Delta {\cal F}_i^{N_0} 
      - \frac{\beta}{2} \{ [{\cal F}_i^p - {\cal F}_i^n] 
                       \pm  [\Delta {\cal F}_i^p - 
                                        \Delta {\cal F}_i^n ]\}\, . 
\label{F2fe}   
\end{equation}  
Here, ``$+$'' and ``$-$'' refer to the neutrino and antineutrino 
structure functions, respectively.  
The last three terms of the right hand side of Eq. \ref{F2fe} 
contain corrections coming from excess neutrons, strange quarks, 
CSV and $s(x)\ne \bar s(x)$. 
Correcting the data for excess neutrons and 
for strange quark contributions  corresponds to  
subtracting the number of events due to the corresponding corrections 
from the left hand side of Eqs. \ref{nneut}    
\begin{eqnarray}
 N^\nu- \sum_{i=2}^3  A_i^\nu  \, (\delta{\cal F}^{\nu}_i)_{n,s} 
 &= &A_2^\nu\,[{\cal F}_2^{N_0}+ 
 (\delta {\cal F}_2^\nu)_{CSV}^{s\bar s}]
   +A_3^\nu\, x[{\cal F}_3^{N_0}+ \delta
({\cal F}_3^\nu)_{CSV}^{s \bar s}] \nonumber\\ 
 N^{\bar\nu}- \sum_{i=2}^3 (-1)^i  
A_i^{\bar\nu} (\delta {\cal F}^{\bar\nu}_i)_{n,s} 
 &= &A_2^{\bar\nu}\,[{\cal F}_2^{N_0}+ (\delta
 {\cal F}_2^{\bar\nu})_{CSV}^{s\bar s}] 
   -A_3^{\bar\nu}\, x[{\cal F}_3^{N_0}+ (\delta
{\cal F}_3^{\bar\nu})_{CSV}^{s\bar s}]\, . 
\label{nneut2}
\end{eqnarray}
In Eq.\ \ref{nneut2}, we have calculated corrections to the structure
functions from excess neutrons and strange quarks, and have used these
to produce the effective number of events on the left hand side of 
Eq.\ \ref{nneut2}.  $(\delta {\cal F}_i)_{n,s}$ and   
$(\delta {\cal F}_i)^{s\bar s}_{CSV}$   
refer to corrections arising from excess neutrons,  
strange quark distributions, 
because of charge symmetry violation and  $s(x)\ne \bar s(x)$, respectively.  
The CCFR Collaboration assumed the validity of 
charge symmetry, and they also took $s(x)=\bar s(x)$ based on the 
results of a next to leading order [NLO] analysis of dimuon 
production in neutrino--induced reactions \cite{CCFRNLO}.  We have
left the correction terms 
coming from CSV and $s(x)\ne\bar s(x)$ on the 
right hand side of Eq. \ref{nneut} as these have been absorbed into the 
extracted structure functions. Under the assumption of charge symmetry 
and $s(x)=\bar s(x)$, Eq.\ \ref{nneut2} simplifies to 
\begin{eqnarray}
 N^\nu- \sum_{i=2}^3  A_i^\nu (\delta {\cal F}^\nu_i)_{n,s}  
 &= &A_2^\nu\,{\cal F}_2^{CCFR,A} +  
   A_3^\nu \, x{\cal F}_3^{CCFR,A}  \nonumber\\ 
 N^{\bar\nu}- \sum_{i=2}^3 (-1)^i A_i^{\bar\nu} 
(\delta {\cal F}^{\bar\nu}_i)_{n,s} 
 &= &A_2^{\bar\nu}\,{\cal F}_2^{CCFR,A} 
   -A_3^{\bar\nu}\, x {\cal F}_3^{CCFR,A}.   
\label{nneut3} 
\end{eqnarray} 
These equations provide a system of two linear equations for the two 
nuclear structure functions ${\cal F}_2^{CCFR,A}$ and 
${\cal F}_3^{CCFR,A}$.  From these structure functions we can 
calculate the structure functions for a ``free'' nucleon target
using the heavy target correction factors described in the 
previous section.  The resulting structure functions 
${\cal F}_i^{CCFR}$ still contain charge symmetry violating 
contributions and terms proportional to $s(x)-\bar s(x)$, as can 
be seen from Eq.\ \ref{eq2}.  To relate the measured structure 
functions, $F_i^{CCFR}$ to the various parton distributions,   
we take the sum and difference of  
the measured number densities in Eqs. \ref{nneut2} and \ref{nneut3}   
(for a fixed energy) and compare the coefficients of $A_i(y)$. In this way,   
we see that the measured structure functions, $F_i^{CCFR}$, can effectively 
be identified with a flux weighted 
average of the neutrino and antineutrino structure 
functions, $F_i^{\nu N_0}$ and $F_i^{\bar\nu N_0}$, and 
correction terms arising from CSV effects  
\begin{eqnarray} 
  F_i^{CCFR} &= &{\cal F}_i^{N_0} + (2\alpha -1) 
   \Delta {\cal F}_i^{N_0} - \frac{\beta}{2} 
     [{\cal F}^p_i-{\cal F}^n_i]_{CSV}   \nonumber\\  
    & - &  \frac{(2\alpha -1)\beta}{2} [\Delta {\cal F}_i^p 
              -\Delta {\cal F}_i^n ]_{CSV}.   
\label{f2ccfr}
\end{eqnarray}  
Here, we defined the relative neutrino flux, $\alpha$, as 
$\alpha\equiv \Phi^\nu/(\Phi^\nu+\Phi^{\bar\nu})$.   
The experimental value of $\alpha$ depends  on the incident neutrino and 
antineutrino energies and is also different for the E744 and E770 
experiments. Because of the kinematical constraint $y<1$, relative 
fluxes at energies $\ge 150$ GeV are relevant for small $x$. 
Here, $\alpha \approx 0.83$ \cite{CCFR} 
so that  $F_2^{CCFR}(x,Q^2)$ can be approximately regarded  
as a neutrino structure function.

The different contributions to $F_2^{CCFR}$  can be expressed 
in terms of the quark distribution functions 
\begin{eqnarray} 
   \frac{1}{2} [{\cal F}^p_2-{\cal F}^n_2]_{CSV} &=& 
   -[{\cal F}^{N_0}_2]_{CSV} 
   =  \frac{x}{2}\,  [\delta u(x) + \delta\bar u(x)
       + \delta d(x) +\delta\bar d(x)]  \nonumber \\ 
   \frac{1}{2} [\Delta {\cal F}_2^p
              -\Delta {\cal F}_2^n]_{CSV} & = & 
          -\frac{x}{2}\, 
       [\delta d(x) -\delta\bar d(x) -\delta u(x) + \delta\bar u(x)] 
    \nonumber \\ \Delta {\cal F}_2^{N_0} &=& - \frac{1}{2} 
    [\Delta {\cal F}_2^p -\Delta {\cal F}_2^n]_{CSV}
  + x[s(x)-\bar s(x)]. 
\label{terms}
\end{eqnarray}  
The second expression in Eq.\ \ref{terms} is obtained by 
subtracting the $F_2$ structure function for neutrinos on
protons from that for antineutrinos on protons; from this is 
subtracted the corresponding term for neutrons.  It depends 
only on charge symmetry violation in the {\it valence} quark
distributions.  The last expression in Eq.\ \ref{terms} is 
obtained by taking the difference between neutrino and antineutrino 
$F_2$ structure functions on an isoscalar system.  It also 
depends on valence quark CSV, and has an additional contribution
from the difference between strange and antistrange 
parton distributions.  The first term in Eq.\ \ref{terms} is obtained by 
averaging the $F_2$ structure functions over neutrino and 
antineutrino reactions, and taking the difference of the $F_2$ 
structure functions measured on proton and neutron targets.  This
quantity is free from strange quark 
effects, and is also sensitive to CSV in the sea-quark 
distributions. 
            
\section{Evidence for Large Charge Symmetry Violation in Parton 
Sea Quark Distributions}

The most likely explanation for the discrepancy in the small-$x$
region of the charge ratio involves either  differences between the 
strange and antistrange quark distributions \cite{Signal,Melni97,JT,HSS},    
or charge symmetry violation. First, we will examine   
the role played by the strange and antistrange quark 
distributions.  Assuming that charge symmetry is exact,  
the strange and antistrange quark distributions are given by a 
linear combination of the structure functions measured in neutrino 
and in muon DIS, as can be seen from Eqs. \ref{eq2}, \ref{f2ccfr} and 
\ref{terms}, 
\begin{equation} 
   \frac{5}{6} F_2^{CCFR}(x,Q^2) -3 F_2^{NMC}(x,Q^2)  
= \frac{1}{2}\, x \, [s(x) + \bar s(x)] 
+\frac{5}{6} (2\alpha -1)\, x \,[s(x)-\bar s(x)]. 
\label{diff} 
\end{equation} 
Under the assumption that $s(x)=\bar s(x)$, this relation could be 
used to extract the strange quark distribution. 
However, as is well known, the strange quark distribution   
obtained in this way is inconsistent with the distribution 
extracted from independent experiments.  

\subsection{Direct Measurement of Sea Quark Distributions}

The strange quark distribution can be determined directly 
from opposite sign dimuon production in deep inelastic neutrino and 
antineutrino 
scattering. To leading order in a charge-changing reaction, the 
incoming neutrino (antineutrino) emits a muon and a virtual $W$ 
boson, which scatters on an $s$ or $d$ ($\bar s$ or $\bar d$)  
quark, producing a charm (anticharm) quark which fragments into a charmed 
hadron. The semi-leptonic decay of the charmed hadron produces an opposite 
sign muon. The CCFR Collaboration performed a LO \cite{CCFRLO}  
and NLO analysis \cite{CCFRNLO} of their dimuon data using    
the neutrino (antineutrino) events to extract the strange 
(antistrange) quark distributions. 
Their result differs substantially from the strange quark distribution 
extracted from Eq.(\ref{diff}), as mentioned above. 

In the dimuon data one extracts the strange and antistrange
quark distributions from the neutrino and antineutrino data separately. 
The analysis performed by the CCFR Collaboration suggests that, while 
there is a difference between the strange and antistrange 
distributions in  LO analysis \cite{CCFRLO}  
they are equal within experimental errors 
in NLO \cite{CCFRNLO}. However, since the number 
of antineutrino events is much smaller than that of the neutrino 
events, the errors of this analysis are inevitably large. 

Since the dimuon experiments are carried out on an iron target,  
shadowing corrections could modify the extracted 
strange quark distribution, and might account 
for some of the discrepancies between the two different  
determinations of the strange quark distributions.  
The CCFR-Collaboration normalized the dimuon cross section 
to the ``single muon'' cross section and argued that 
the heavy target correction should cancel in the ratio. 
However, the charm producing part of the structure function  
$F_2^{cp}(x,Q^2)$ could be shadowed differently  
from  the non-charm producing part $F_2^{ncp}(x,Q^2)$,  
unless charm threshold effects cancel in the shadowing ratio. 
This could be the case, because vector mesons with higher masses 
are involved in the charm producing part, and because 
charm production threshold effects have to be taken into 
account in the Pomeron component as well. 

\begin{figure}
\epsfig{figure=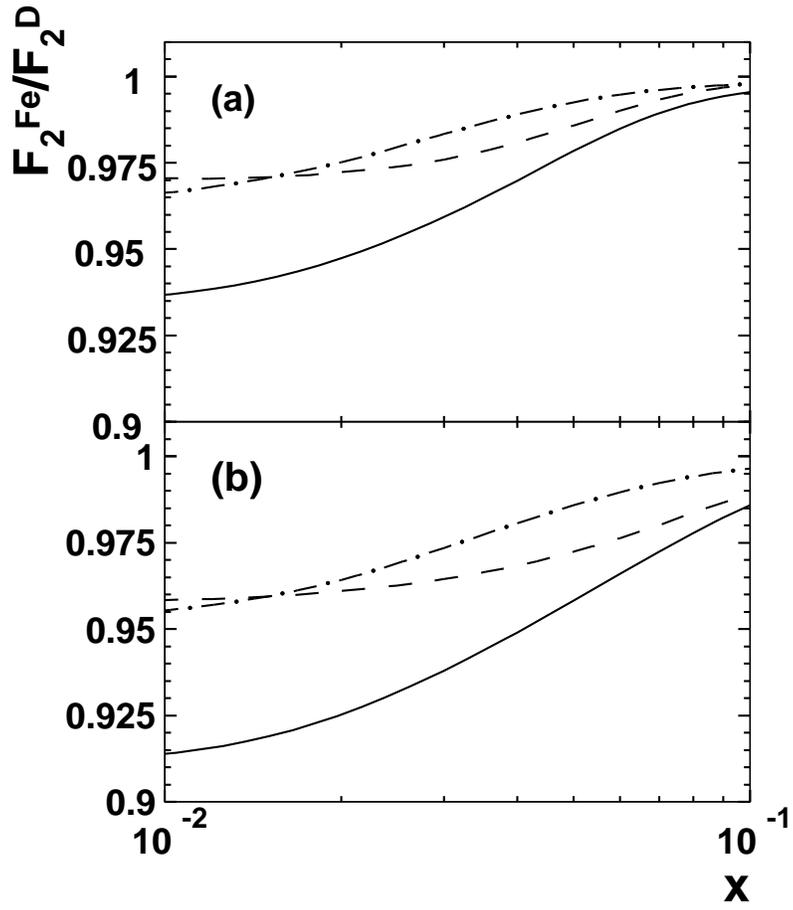,height=14.cm}
\caption{Shadowing corrections in  
    the (a) charm  and (b) non-charm producing parts of the 
    neutrino structure function, as a function 
    of $x$, for a fixed $Q^2=5$ GeV$^2$. The dashed (dash-dotted)  
    lines stand for VMD (Pomeron) contributions. The solid lines 
    represent the total shadowing.}
\label{fig3}
\end{figure}

We calculated the shadowing ratio, $R\equiv 
F_2^{\nu A}(x,Q^2)/F_2^{\nu D}(x,Q^2)$,  
between the structure functions on a heavy target 
and on a deuteron target for both the 
charm and non-charm producing part of the structure 
function. We took charm production threshold effects into account 
in the Pomeron component through the 
slow rescaling mechanism  
by replacing $x_{I\!P}$, which is the momentum fraction of the 
Pomeron carried by the struck quark, by 
$\xi_{I\!P}=x_{I\!P}(1+\frac{m_c^2}{Q^2})$. 
Here $m_c$ is the mass of the charm quark. In the VMD 
component of $F_2^{cp}(x,Q^2)$ we included the vector mesons 
$D^{*+}(2010)$, $D^{*+}_s(2110)$ and the axial vector partner  
$D^{*+}_{As}(2535)$ of $D^{*+}_s$ \cite{Data}, which describe the lightest 
coherent states of the $c\bar d$ and $c\bar s$ fluctuations of the 
$W^+$-boson. 
They have the same coupling to 
$W^+$ as $\rho^+$ and $a_1^+$ but have much heavier masses.  
(The $c\bar d$ fluctuations are suppressed by sin$^2\Theta_c$.)   
Because of the larger mass of the charmed vector mesons 
($\sim 2.5$ GeV), we applied a cut at $M^2_X\ge 6.3$ GeV$^2$ 
in the diffractively 
produced invariant mass of the Pomeron component. 
This is to be compared with $M^2_X\ge 1.5$ GeV$^2$ in the non-charm producing 
part of the structure function (these cuts are necessary to avoid 
double counting.)  Because of the {\it light} quark component 
of the D-mesons, we expect that the D-meson-nucleon total 
cross sections are comparable to the corresponding cross sections 
of lighter mesons with the same light quark content. We 
use $\sigma_{D^*N}\approx \sigma_{\rho N}$ and 
$\sigma_{D^*_sN}\approx \sigma_{\phi N}$. 
The calculated ratios, 
$R=F_2^A(x,Q^2)/F_2^D(x,Q^2)$,   
are shown in Fig. \ref{fig3} for $Q^2=5$ GeV$^2$.   
Here,  $F_2^A=F_2^D+\delta F_2^{(V)}+F_2^{I\!P}$ and 
$\delta F_2^{(V)}$ and $F_2^{(I\!P)}$, the shadowing 
corrections to the structure functions 
due to vector mesons and Pomeron-exchange, 
respectively, are calculated in 
the two phase model.  Since the pion component is negligible for 
$Q^2=5$ GeV$^2$, we did not include it.  

There is  no substantial difference 
in shadowing between the charm producing ($cp$) and non-charm  
producing ($ncp$) parts. (The difference is about $2\%$ in the 
small $x$-region).  
Note that the shadowing correction in 
$F_2^{cp}(x,Q^2)$ decreases faster with increasing 
$x$, because the larger masses of the charmed 
vector mesons, $m_V$,  enter in the coherence condition   
$\tau =\frac{1}{Mx}(1+\frac{m_V^2}{Q^2})^{-1}$ 
($\tau$ is the lifetime of the quark antiquark fluctuation, 
and $M$ is the nucleon mass),  
compared with the smaller masses of the $\rho$ and $a_1$.  
Our results justify the assumption that shadowing corrections 
approximately cancel  in the ratio of dimuon and single muon cross sections. 

\subsection{Estimate of Parton CSV Contribution}

It would appear that a likely explanation for the deviation of
the charge ratio of Eq.\ \ref{rc} from one is due to differences
between the strange and antistrange quark densities. 
To test this hypothesis, we combined the 
data in dimuon production, averaged over both neutrino and 
antineutrino events, with the difference between the 
structure functions in neutrino and charged lepton scattering 
(Eq.(\ref{diff})).   
In combining the neutrino and antineutrino events, one measures a 
flux-weighted average of the strange and antistrange quark distributions. 
If we define $\alpha^\prime= N_\nu/(N_\nu+N_{\bar\nu})$, where 
$N_\nu =5,030 $, $N_{\bar\nu}=1,060$ ($\alpha^\prime \approx 0.83$)  
are the number of neutrino and antineutrino 
events of the dimuon production experiment \cite{CCFRNLO}, 
we have for the measured distribution $x s(x)^{\mu\mu}$  
\begin{equation} 
 x s^{\mu\mu}(x) = \frac{1}{2}\, x \,[s(x) + \bar s(x)] + \frac{1}{2}  
       (2\alpha^\prime -1 )\, x\, [s(x) - \bar s(x)].  
\label{s2}
\end{equation} 
Now, this  equation together with Eq.(\ref{diff}) 
forms a pair of linear equations which can be solved for 
$\frac{1}{2} x [s(x)+\bar s(x)]$ and  $\frac{1}{2} x [s(x)-\bar s(x)]$.   
In this way we can also test the compatibility of the two 
experiments.  In addition we have the sum rule that the nucleon 
contains no net strangeness, 
\begin{equation}
  \int_0^1 [s(x) - \overline{s}(x)]\, dx = 0
\label{smrule}
\end{equation} 
In the following expressions, we have not enforced the sum rule
requirement on the antistrange quark distributions. 

Compatibility of the two experiments requires that physically 
acceptable solutions  for 
$\frac{1}{2} x [s(x)+\bar s(x)]$ and  $\frac{1}{2} x [s(x)-\bar s(x)]$,  
satisfying both Eq. \ref{diff} and Eq. \ref{s2}, can be found. 
Using the experimental values $\alpha = \alpha^\prime \approx 0.83$, 
we can write $x [s(x)-\bar s(x)]=\Delta(x)/\delta$, 
where $\Delta(x) =\frac{5}{6}F_2^{CCFR}(x) -3 F_2^{NMC}(x)-
s^{\mu\mu}(x)$, and $\delta = (2\alpha -1)/3 \approx 0.22$.  
Consequently even rather small values for $\Delta(x)$ can lead to 
large differences between $s$ and $\bar s$.  Note that the value 
of the relative neutrino flux, $\alpha$, depends on the incident 
neutrino energy. While $\alpha\approx 0.83$ for small $x$, $\alpha$  
is somewhat smaller for higher $x$-values. However, smaller 
$\alpha$ would lead to an even smaller $\delta$ and  
would require even larger differences between $s$ and $\bar s$.   

\begin{figure}
\epsfig{figure=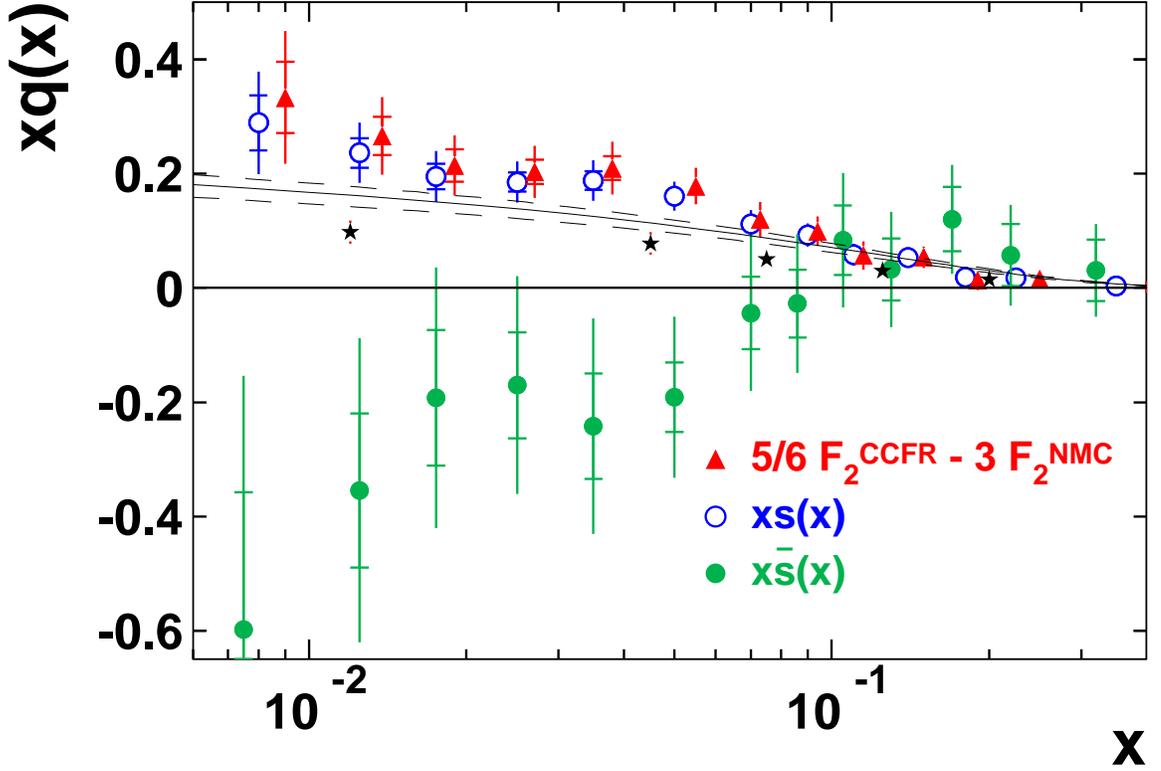,height=14.cm}
\caption{The strange quark distribution $x\,s(x)$ (open circles) and 
      antistrange distribution $x\,\bar{s}(x)$ (solid circles) 
      extracted from the CCFR and NMC structure functions.  
      The difference between the CCFR neutrino and NMC muon 
      structure functions $\frac{5}{6} F_2^{CCFR}-3F_2^{NMC}$ 
      (see Eq.\ \protect\ref{diff}) is shown as solid 
      triangles. The strange quark distribution extracted 
      by  CCFR in a LO-analysis \protect\cite{CCFRLO} is shown as solid 
      stars, while that from a NLO-analysis \protect\cite{CCFRNLO} is 
      represented by the solid line, with a band indicating  $\pm 1\sigma$ 
      uncertainty in the distribution. Statistical and systematic errors 
      are added in quadrature.}
\label{fig4}
\end{figure}

\begin{figure}
\epsfig{figure=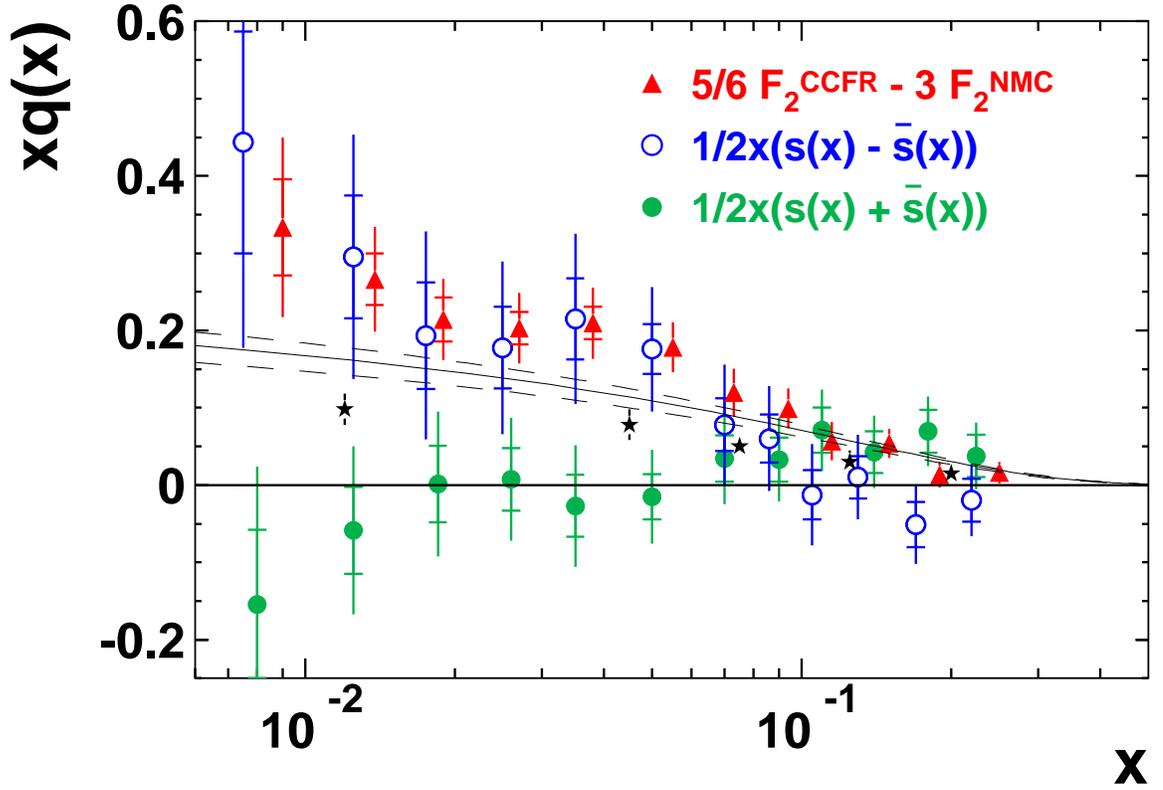,height=14.cm}
\caption{$\frac{1}{2}x[s(x)+\bar s(x)]$ (solid circles) and 
          $\frac{1}{2}x[s(x)-\bar s(x)]$  (open circles)  
      as extracted from the CCFR and NMC structure functions 
      and from the dimuon production data. See Fig.\ 
      \protect\ref{fig4} for the 
      definition of the other quantities. 
      The strange quark distribution extracted by the CCFR 
      Collaboration is shown as a solid line with a band 
      indicating $\pm 1\sigma$ uncertainty.  
      Statistical and systematic errors are added in
      quadrature. }
\label{fig5}
\end{figure}

\begin{figure}
\epsfig{figure=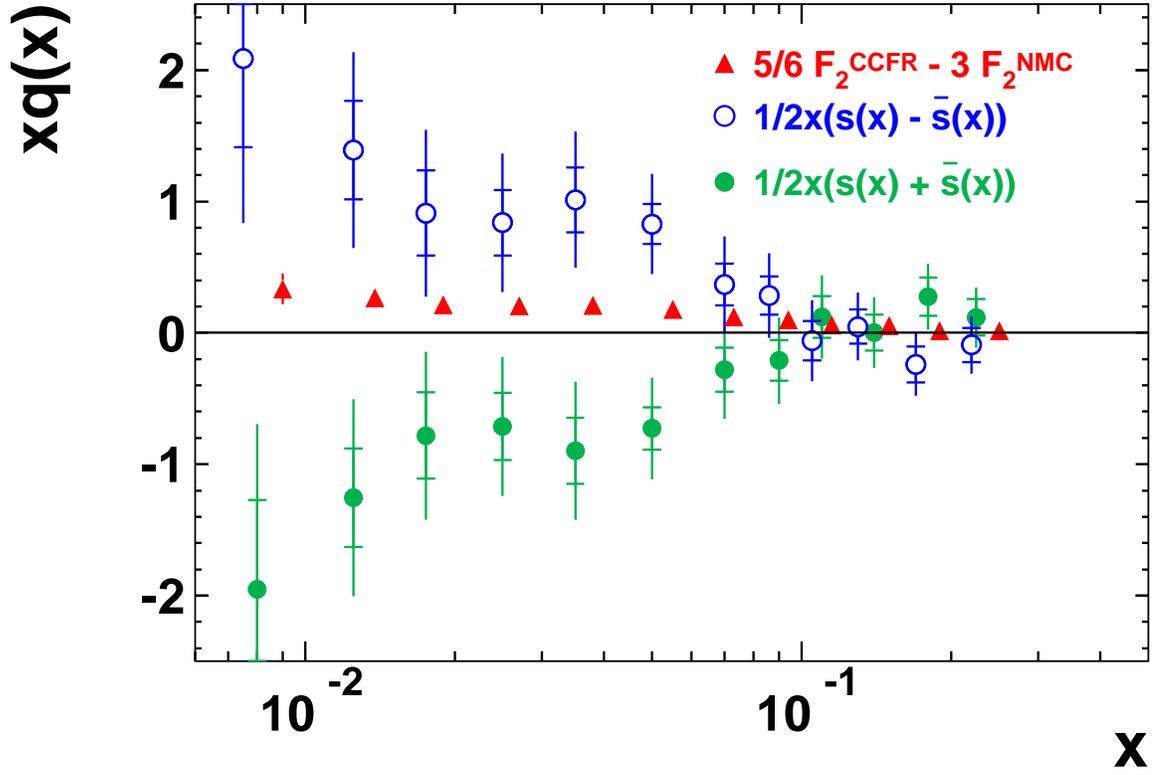,height=14.cm}
\caption{$\frac{1}{2}x[s(x)+\bar s(x)]$ (solid circles) and
          $\frac{1}{2}x[s(x)-\bar s(x)]$  (open circles)
      as extracted from the CCFR and NMC structure functions
      and from the dimuon production data using  $\alpha^\prime =1$.}
\label{fig6}
\end{figure}

In Fig.\ \ref{fig4} we show the results obtained for $x s(x)$ (open 
circles) and $x \bar s(x)$ (solid circles) by solving the resulting
linear equations, Eqs.\ \ref{diff} and \ref{s2} 
using the values $\alpha =\alpha^\prime =0.83$.  The results are 
completely unphysical, since the antistrange 
quark distribution is negative, which is not possible since the 
distribution is related to a probability.  In Fig.\ \ref{fig5} we 
show the corresponding results
for the linear combinations $\frac{1}{2}x[s(x)+\bar s(x)]$ (solid
circles) and $\frac{1}{2}x[s(x)-\bar s(x)]$ (open circles).  
The unphysical nature of the solution is demonstrated by the fact
that $\frac{1}{2}x[s(x)-\bar s(x)]$ is larger than 
$\frac{1}{2}x[s(x)+\bar s(x)]$.   
We also solved the equations using 
the values $\alpha = 0.83$ and $\alpha^\prime =1$  
which corresponds to
using  a subsample of the di-muon data containing only
neutrino events. In this case, even the {\it sum} of the strange and 
anti strange distributions is negative.  This is shown in Fig. \ref{fig6}.  
 
Thus, our analysis strongly suggests that the discrepancy between 
$F_2^{CCFR}(x,Q^2)$ and $F_2^{NMC}(x,Q^2)$ cannot be completely 
attributed to differences between the strange and antistrange 
quark distributions. In other words, assuming parton charge
symmetry the two experiments are incompatible with each other,  
even if the antistrange quark distribution is allowed 
to be different from the strange distribution. (Note, that  
absolutely {\it no} restrictions were placed on the
antistrange quark distribution, aside from the condition 
that since it represents a 
probability density, it must be non negative.) 
We stress that  our conclusion is quite different from that 
of Brodsky and Ma \cite{Brodsky},  
who  suggested that allowing 
$s(x)\ne \bar s(x)$ could account for the difference between the two 
determinations of the strange quark distribution.  
However, they treated the  
CCFR structure functions 
as an average beween the neutrino and the antineutrino structure 
function which corresponds to  setting $\alpha =0.5$.    

At this point there are two possibilities to explain the 
low-$x$ discrepancy observed between the CCFR neutrino and
the NMC muon structure functions.  Either one of the experimental 
structure functions (or the strange quark distributions) is incorrect 
at low $x$, or parton charge symmetry is violated in this
region, since we have shown that neither neutrino shadowing
corrections nor an inequality between strange and antistrange
quark distributions can explain this experimental anomaly.  
If we include the possibility of parton CSV, then we 
can combine the dimuon data for the strange quark distribution,  
Eq.\ \ref{s2}, with the relation between neutrino and muon
structure functions, Eq.\ \ref{diff}, to obtain the relation 
\begin{eqnarray} 
   \frac{5}{6} F_2^{CCFR}(x,Q^2) &-&3 F_2^{NMC}(x,Q^2)
 -x s^{\mu\mu}(x) = \frac{(2\alpha -1)\,x}{3}
  [s(x) -\bar{s}(x)] \nonumber \\ 
  &+& \frac{(3-5\beta)\,x}{12} \, [\delta d(x) + \delta\bar d(x)] - 
 \frac{(3+5\beta)\,x}{12} \, [\delta u(x) + \delta\bar u(x)] \nonumber \\ 
 &-& \frac{5(1+\beta)(2\alpha-1)\,x}{12}[\delta u_v(x) -\delta d_v(x)]    
\label{csv1} 
\end{eqnarray} 
In Eq.\ \ref{csv1} we have used the experimental value 
$\alpha = \alpha^\prime$, and we have defined the valence quark CSV
terms $\delta q_v(x)\equiv \delta q(x) -\delta \bar q(x)$.  We have 
neglected the effects of possible CSV on the extraction of 
$s(x)$ from the dimuon data, or on the identification of the structure
functions from the neutrino data. This will be discussed below. 
Since the discrepancy between CCFR and NMC data lies 
primarily in the very small $x$-region, where the valence quark
distribution is much smaller than the sea quark, the charge 
symmetry violation should be predominantly 
in the sea quark distributions. If we set 
$\delta q_v(x) \approx 0$ in 
this region, Eq.(\ref{csv1}) can be written as 
\begin{eqnarray} 
   \frac{5}{6} F_2^{CCFR}(x,Q^2) &-& 3 F_2^{NMC}(x,Q^2)
 - x s^{\mu\mu}(x) \approx \frac{(2\alpha -1)\,x}{3}[s(x) 
 -\bar{s}(x)] \nonumber\\
  &+& \frac{x}{2} \, [\delta \bar d(x) -\delta\bar u(x)]
  - \frac{5\beta\,x}{6} \, [\delta \bar d(x) +\delta\bar u(x)]~~. 
\label{csv2}
\end{eqnarray} 
Since $\beta \approx 0.06$ is quite small, CSV arising from the 
non-isoscalar nature of the iron target can be neglected, so in
the following we neglect the last term of Eq.\ \ref{csv2}.  
  
\begin{figure}
\epsfig{figure=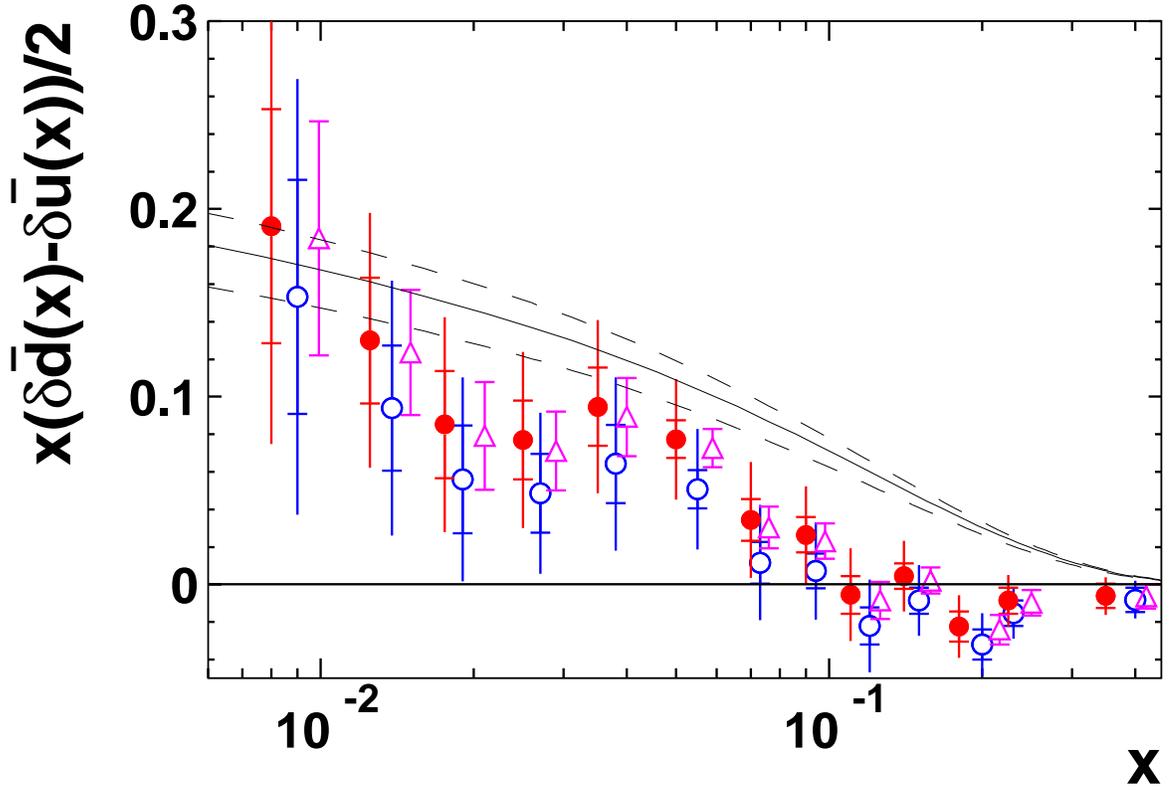,height=14.cm}
\caption{
Charge symmetry violating distributions $x(\delta\bar{d}(x) - 
	\delta\bar{u}(x))/2$ extracted
      from the CCFR and NMC structure function data and
      the CCFR dimuon production data under the assumption
      that $s(x)=\bar s(x)$ (solid circles) and
      $\bar s(x)\approx 0$ (open circles) for
        $\alpha^\prime =0.83$, and $s(x)=\bar s(x)$ (solid circles) and
      $\bar s(x)\approx 0$ (open triangles) for $\alpha^\prime =1$.
      (For the latter only statistical errors are shown.)
       $xs(x)$ at $Q^2=4$ GeV$^2$ obtained by the CCFR
       Collaboration in a NLO analysis
     \protect\cite{CCFRNLO} is shown for comparison (solid curve, with
     $1\sigma$ error band).}
\label{fig7}
\end{figure}

Using the experimental data we find that 
the left hand side of Eq.\ \ref{csv2} is positive.  Consequently, 
the smallest value for charge symmetry violation will be obtained
if we set $\bar{s}(x) = 0$ \cite{smrul}.  In Fig.\ref{fig7} we show the 
magnitude of charge symmetry violation needed to satisfy the
experimental values in Eq.\ \ref{csv2}.  The open circles 
are obtained if we set $\bar{s}(x) = 0$, and the solid circles 
result from setting $\bar{s}(x) = s(x)$.   
If we use only the neutrino induced di-muon events, 
(i.e we set $\alpha^\prime =1$),      
the coefficient of $x[s(x)-\bar s(x)]$, $(5\alpha -3\alpha^\prime -1)/3$,  
is still positive but  smaller in  
magnitude. Consequently, the influence of the  uncertainty in $\bar s(x)$ 
on the extracted CSV is smaller.   
This is shown as open triangles in Fig. \ref{fig7}.    

\begin{figure}
\epsfig{figure=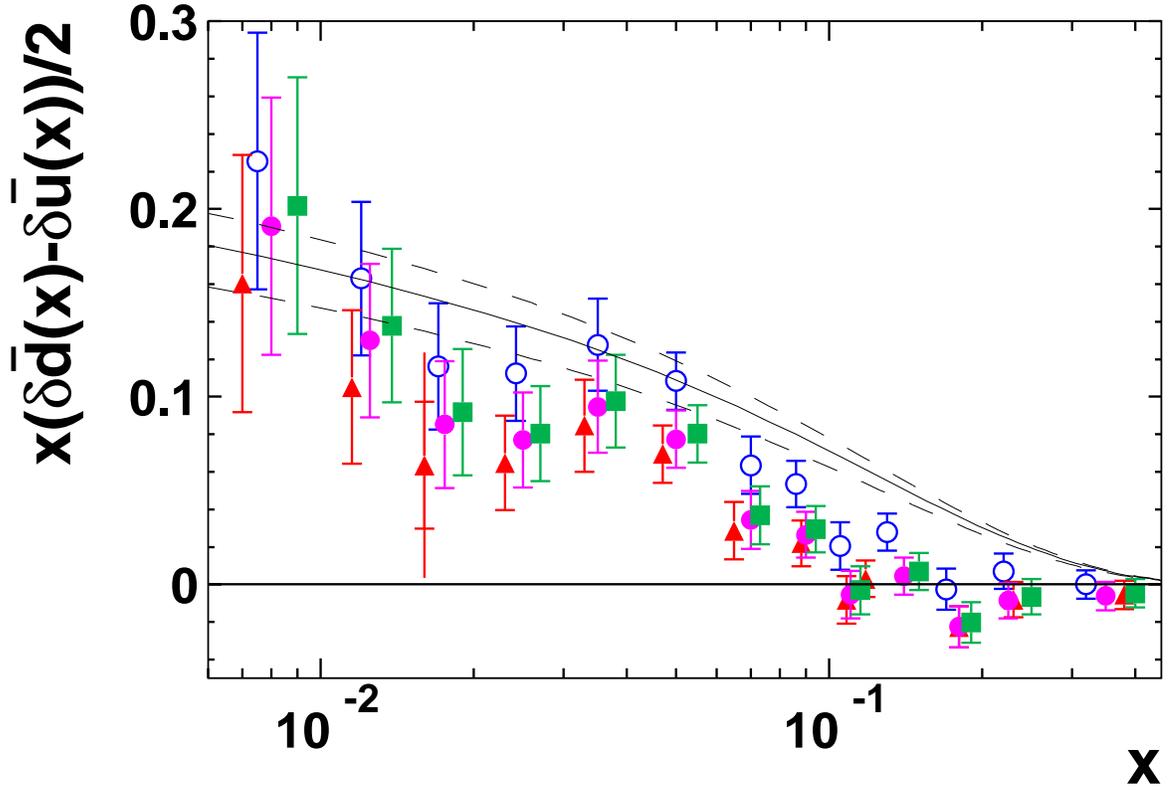,height=14.cm}
\caption{Uncertainty in the extracted parton CSV term $x(\delta\bar{d}(x) - 
	\delta\bar{u}(x))/2$ due to the parametrization used
       for the dimuon data on charge symmetry violation.
       Open circles: LO CCFR distribution, solid circles:
       CTEQ4L parton distribution \protect\cite{Lai}; solid rectangles: 
       CTEQ4D parton distribution; solid triangles: NLO CCFR distribution.
       Here, except for the most ``critical'' point,
       only statistical errors are shown.}
\label{fig8}
\end{figure}

In obtaining these results,   
both the structure functions and the strange quark distribution  
have been integrated over the overlapping kinematic regions and 
we used the CTEQ4L parametrization for $s^{\mu\mu}$ 
\cite{Lai}.      
In Fig.\ref{fig8} we show the sensitivity of extracted CSV  
to the parametrization used for $s^{\mu\mu}$. 
The uncertainty due to different parametrizations  
has been partly  taken into account since the calculated errors already 
include the uncertainty of the dimuon measurement and most of the 
parametrizations lie within the experimental  
errors of the dimuon data (except for LO-CCFR $s(x)$).  
We note that  the magnitude of the observed charge symmetry 
violation in the sea quark distributions is  
independent of whether we use a pure neutrino or antineutrino 
structure function or a linear combination of neutrino and 
antineutrino structure functions.  This is quite different from strange 
and antistrange quark effects which are sensitive to the relative  
weighting of neutrino and antineutrino events in the data sample.  
Thus, effects due to CSV are independent of the precise value of the 
relative neutrino flux, $\alpha$. 
  
The CSV effect required 
to account for the NMC-CCFR discrepancy is extraordinarily large.  
It is roughly the same size as the strange quark distribution at
small $x$ (compare the open circles in Fig.\ \ref{fig7} with the solid line in 
Fig. \ref{fig7}).  The charge symmetry violation necessary to 
provide agreement with the experimental data is about 25\% of the 
light sea quark distributions for $x < 0.1$.   
The level of CSV required is 
two to three orders of magnitude larger than the theoretical estimates 
of charge symmetry violation \cite{Ben98,Lond,Sather,Ben97}.    
Note that, if $\bar s(x) < s(x)$ in this region,   
as suggested in Ref.\cite{Brodsky},  
we would need an even larger CSV to account for the 
CCFR-NMC discrepancy.  

Theoretical considerations suggest that $\delta\bar d(x) \approx 
-\delta\bar u(x)$ \cite{Ben98,Lond}. In fact, since 
charge symmetry violation seems to be surprisingly large, it is 
reasonable to assume that these distributions have opposite 
signs.  
We note that with this sign CSV effects also require large flavor
symmetry violation.  One might ask whether such large CSV
effects would be seen in other experiments.  For example, CSV 
in the nucleon sea could contribute to the observed violation of the 
Gottfried sum rule 
\cite{Ma1,Lond,Ben98} and could explain the  Fermilab Drell-Yan experiment 
\cite{Ma1}.  This will be discussed in section IV.

Clearly, CSV effects of this magnitude need further experimental 
verification. The NuTeV-experiment at Fermilab \cite{NuTeV} is able to 
operate either with pure neutrino or pure antineutrino 
beams. The extracted structure functions can be used to build 
different linear combinations, proportional to 
various combinations of the $\delta\bar q$'s and $s$-$\bar s$. 
This will be useful to separate CSV from $s$-$\bar s$ effects. 

At small $x$, our results can be summarized by 
\begin{eqnarray}
\delta \bar{d}(x) - \delta \bar{u}(x) &\approx & {1\over 2} (s(x) 
  + \bar{s}(x) ) \approx {1\over 4} \left({\bar{u}(x) + \bar{d}(x) 
  \over 2} \right) \nonumber \\ 
  \delta \bar{d}(x) + \delta \bar{u}(x) &\approx & 0 .  
\label{csvappr}
\end{eqnarray}  
From Eq.\ \ref{eq2} we note that such a CSV effect would have 
little or no effect on the the $F_2$ structure functions of  
isoscalar targets, for either neutrinos or antineutrinos.  The major 
effect for isoscalar targets would be a 
significant positive contribution to $F_3^{\nu N_0}(x,Q^2)$ at small $x$, 
and an equally large negative contribution to 
$F_3^{\bar\nu N_0}(x,Q^2)$.    

However, if CSV effects of this magnitude are really present
at small $x$, then we should include charge symmetry violating
amplitudes in parton phenomenology from the outset, and re-analyze 
the extraction of
all parton distributions.  Given the experimental values 
$\kappa = 2S/(U+D) \approx 0.5$, where $S$, $U$ and $D$ are the 
probabilities for strange, up and down quarks averaged over $x$, 
and the size of CSV effects suggested by the preceeding analysis,   
we would predict that at small $x$, $\bar{d}^n(x) \approx 
1.25\,\bar{u}^p(x)$ and $\bar{u}^n(x) \approx 0.75\,\bar{d}^p(x)$.

\section{Effects of Parton CSV on Other Observables} 

If there is substantial CSV, it should also effect  
other observables. In the following we review the effects which such
large CSV terms might have on three quantities; first, the recent 
search for parton ``flavor symmetry violation'' [FSV] by the Fermilab 
Drell-Yan experiment E866; second, the extraction of the strange 
quark distribution; and third, experimental determination of the
Weinberg angle $\sin^2(\theta_W)$. 

\subsection{Flavor Symmetry Violation in the Proton Sea}

The results of the recent Fermilab Drell-Yan experiment \cite{E866} and 
the comparison of the proton and neutron structure functions  
measured by the NMC Collaboration \cite{NMCfsv} indicate substantial 
flavor symmetry violation. However, both experimental observations  
could be 
attributed to charge symmetry violation, as pointed out by Ma \cite{Ma1} 
(see also \cite{Steffens}).   
Furthermore, both CSV and FSV could be present, as suggested by our analysis 
of the CCFR-NMC discrepancy. Therefore, it is 
important to examine the effects of CSV  
on the interpretation of the Fermilab and NMC experiments. 

First, we discuss the Drell-Yan experiment which measures 
the ratio of the dimuon cross sections from proton-deuteron 
and proton-proton scattering. Since CSV is significant in the small 
$x$ region, it is a reasonable first approximation 
to keep only the contributions to the Drell-Yan cross sections 
which come from the annihilation of 
quarks of the projectile and antiquarks of the target \cite{seacomm}.  
In this approximation, the ratio $R\equiv \sigma^{pD}/(2\sigma^{pp})$ 
is given by 
\begin{equation}
\frac{\sigma^{pD}}{2\sigma^{pp}}
\approx \frac{[1+\frac{\bar d_2}{\bar u_2} -
  \frac{\delta\bar d_2}{\bar u_2}] +\frac{R_1}{4}[1+\frac{\bar d_2}{\bar u_2}
-\frac{\delta\bar u_2}{\bar u_2}]}{2\left( 1+\frac{R_1}{4}\frac{\bar d_2}
{\bar u_2}\right)} . 
\end{equation}
Here, we introduced the notation $q_{j}\equiv q(x_j)$ for the quark 
distributions ($x_1$ is the projectile $x$ value and $x_2$ refers to
the target), and $R_1\equiv \frac{d_1}{u_1}$.  
    
For large $x_F$, which corresponds to large $x_1$, the quantity 
$R_1$ is small; if we ignore it, we have the approximate result 
\begin{equation}
  R =\frac{\sigma^{pD}}{2\sigma^{pp}}
   \approx \frac{1}{2} \{ 1 + \frac{(\bar d_2 -
\delta\bar d_2)}{\bar u_2} \} . 
\label{r}
\end{equation} 
If charge symmetry is valid and $\bar d_2 = \bar u_2$, then we 
would have $R=1$.  The experimental values give $R > 1$ at small 
$x_2$; from Eq.\ \ref{r}, this could be satisfied if either 
$\bar d_2 > \bar u_2$ or $\delta\bar d_2$ was large and negative.  
However, the value of $\delta\bar d(x)$ extracted from the existing
neutrino and muon experiments, as discussed in the preceding 
section, was large and positive at small $x$.  The enhancement is on 
the order of 
$25\%$ in the small $x$ region where CSV could be important. 

\begin{figure}
\epsfig{figure=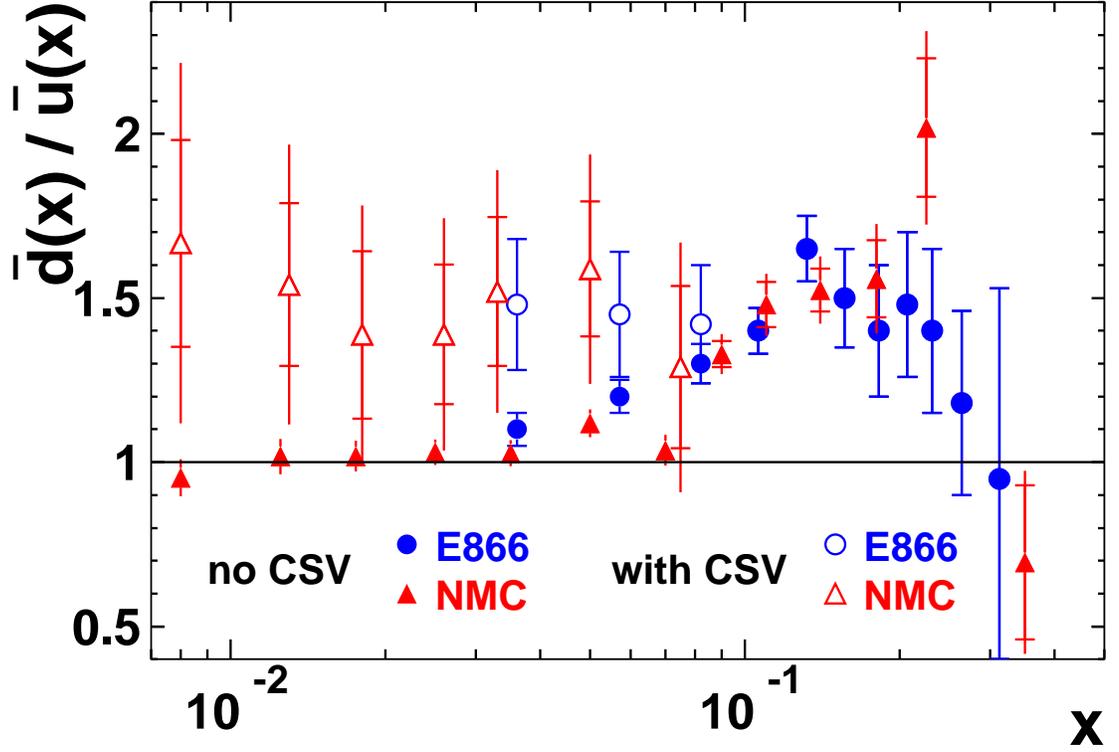,height=14.cm}
\caption{Solid circles: the ratio $\bar{d}(x)/\bar{u}(x)$ vs.\ $x$, 
     extracted from the Drell-Yan data of FNAL experiment E866 
     \protect\cite{E866} assuming the validity of charge symmetry. 
     If CS is violated this ratio corresponds to 
     $(\bar{d}(x)-\delta \bar{d}(x))/\bar{u}(x)$. The result obtained by 
     correcting for CSV is shown as open circles. 
     The ratio $\bar{d}(x)/\bar{u}(x)$ extracted from the difference of  
     proton and deuteron structure functions measured by the NMC group 
     \protect\cite{NMCfsv} is shown as solid and 
     open triangles, without and with CSV, respectively.} 
\label{fig9}
\end{figure}

In Fig. \ref{fig9} the solid circles show the ratio 
$\bar{d}(x)/\bar{u}(x)$ extracted from the Drell-Yan experiment 
if we assume the validity of charge symmetry.  The open circles 
in Fig. \ref{fig9} show
the result for $\bar{d}(x)/\bar{u}(x)$ if we include 
the CSV term which was extracted from the CCFR--NMC data (this
is shown in Fig.\ \ref{fig7}). 
Inclusion of parton charge symmetry violation 
suggested by the CCFR-NMC discrepancy plays an important role in the 
extraction of the FSV ratio $\bar d/\bar u$, in the region $x<0.1$.   
The flavor symmetry violation in the sea has to be substantially 
larger to overcome the CSV term which goes in the opposite direction.  
In particular, the ratio $\bar d(x)/\bar u(x)$ does not approach 1 for 
small $x$ values.  

We can invert the extracted ratio to obtain the difference 
$[(\bar d -\delta\bar d) -\bar u]$ 
\begin{equation}
 (\bar d -\delta\bar d) -\bar u =\frac{(\bar d - \delta\bar d)/\bar u -1}
{(\bar d - \delta\bar d)/\bar u+1}
  [(\bar d - \delta\bar d)+ \bar u] . 
\label{inv}
\end{equation}
As a rough approximation, we could neglect $\delta \bar d$ in the 
sum on the right hand side of Eq. \ref{inv} 
and keep it in the difference between $\bar d$ and $\bar u$ on the left 
hand side. For $\bar u + \bar d$ one could use a parametrization. This is 
exactly the way that $\bar d -\bar u$ has been extracted from the Drell-Yan 
data, so that in fact the extracted quantity corresponds to 
$(\bar d -\delta\bar d) -\bar u$ if CSV is present. 

The difference, $\bar d -\bar u$, can also be extracted 
from the difference between the proton and neutron structure 
functions measured by the NMC Collaboration \cite{NMCfsv} using 
muon deep inelastic scattering.  In this case we have 
\begin{equation}
   \frac{1}{2}(u_v(x)-d_v(x))-\frac{3}{2x}(F_2^p(x)-F_2^n(x)) =
   (\bar d(x) -\bar u(x) ) -\frac{2}{3}(\delta d(x) +\delta \bar d(x))
  -\frac{1}{6} (\delta u(x) +\delta\bar u(x)). 
\end{equation}
We can make the approximations $\delta q(x) \approx \delta\bar{q}(x)$
and $\delta\bar{d}(x) \approx -\delta\bar{u}(x)$, (the latter may not be
a good approximation since we have FSV), and obtain  
\begin{equation}
   \frac{1}{2}(u_v(x)-d_v(x))-\frac{3}{2x}(F_2^p(x)-F_2^n)(x) \approx 
   [(\bar{d}(x) -\delta\bar{d}(x))-\bar{u}(x)].
\label{nmcdiff} 
\end{equation}
Comparing this with Eq. \ref{inv} we see that, in a first approximation, 
the quantities extracted from the two experiments are the same even if 
both CSV and FSV are present. However, if CSV is present, the 
term $\delta\bar d$ has to be subtracted from the measured 
quantity to obtain the difference $\bar d -\bar u$. 

We inverted Eq. \ref{nmcdiff} by dividing both sides by   
$\bar d-\delta \bar d +\bar u \equiv \bar u (r_2+1)$, approximating 
$\bar d-\delta \bar d +\bar u$ on the left hand side of Eq. \ref{nmcdiff} 
by a parametrization of  $\bar d + \bar u$ and solving for $r_2 = 
\bar{d}(x_2)/\bar{u}(x_2)$. 
The structure functions and the parton distribution 
are integrated for each data point 
over  the same $Q^2$ regions as in the analysis of the charge ratio. 
The result is shown in Fig. \ref{fig9} as solid triangles.  
If we subtract the contribution of CSV from the ratio $r_2$ we obtain 
the result shown as open triangles in Fig. \ref{fig9}.  
We see  that charge symmetry violation, as  
suggested by the CCFR-NMC discrepancy, considerably enhances  
the FSV ratio $\bar d/\bar u$ in the region $x<0.1$.   

It is interesting to investigate the influence of CSV on the
Gottfried sum rule. If both CSV and FSV are present
the Gottfried sum rule can be expressed as
\begin{equation}
   S_G=\frac{1}{3} - \frac{2}{3} \int_0^1 dx [\bar d(x) - \bar u(x)]
          +\frac{2}{9} \int_0^1 dx [4\delta\bar d(x) +\delta\bar u(x)].
\end{equation}
Now, if $\delta\bar d(x)\approx -\delta\bar u(x)$
we have
\begin{equation}
   S_G=\frac{1}{3} - \frac{2}{3} \int_0^1 dx \{[\bar d(x) -
  \delta\bar d(x)] - \bar u(x)\}, 
\end{equation}
so that, although the CSV suggested by the CCFR experiment does influence 
the magnitude of the extracted 
FSV, it does not change the experimental value of the Gottfried sum rule
since the extracted quantities appear in exactly the same form in the 
Gottfried sum rule as in the Drell-Yan and NMC experiments. 

\subsection{Extraction of Strange Quark Distributions}

The differential cross section for the production of opposite-sign 
dimuons, for neutrino and antineutrino deep inelastic 
scattering from an isoscalar target, are proportional to the quark 
distributions, the 
CKM-matrix elements, the fragmentation function $D(z)$ of the struck 
quark and the weighted average of the 
semi-leptonic branching ratios of the charmed hadrons 
$B_c$    
\begin{equation} 
  \frac{d\sigma (\nu N_0\rightarrow \mu^-\mu^+ X)}
  {d\xi dy dz} \sim \{
  [u(\xi ) + d (\xi ) -\delta u (\xi )] |V_{cd}|^2 
  + 2 s(\xi ) |V_{cs}|^2 \} D(z) B_c(c \rightarrow \mu^+ X). 
\label{dimuon}
\end{equation} 
For antineutrino scattering the quark distributions should be 
replaced by the corresponding antiquark distributions.  
In this equation, $\xi$ is the rescaling variable 
defined by $\xi=x(1+\frac{m^2_c}{Q^2})$, with $m_c$ 
the mass of the produced charm quark.  
The CCFR-Collaboration used
this expression together with 
the parametrization of the quark distributions extracted 
from their structure function data to determine the strange quark 
distributions.  
The strange quark component of the quark sea was allowed 
to have a different magnitude and shape from the 
non-strange component. These two degrees of freedom 
were parametrized by two free parameters $\kappa$ and $\alpha$, 
respectively. 
Further, they treated $B_c$ and the mass of the charm quark, $m_c$,  
as free parameters and  performed a $\chi^2$ minimization 
to find the four free parameters by fitting to 
distributions of the measured number densities.
We note first, that, provided $\delta \bar u=-\delta \bar d$ 
(see Eqs.(\ref{eq2}) and (\ref{csvappr})), charge symmetry violation 
does not effect the extraction of the non-strange parton distributions 
from the structure function data for small $x$-values.
For an isoscalar target, these distributions 
can be determined quite accurately, even if charge symmetry is broken 
in the manner given by Eq.\ \ref{csvappr}. 
However, in extracting the strange quark distribution, 
charge symmetry violation plays an important role.  
The distribution extracted by the CCFR-Collaboration is {\it  
not} the strange quark distribution, but a linear combination  
of the true strange quark distribution and 
the term in Eq.(\ref{dimuon}) coming from CSV. Hence, 
the distribution measured in the experiment,  
$s^{CCFR}(x)$,  is related to the ``true'' strange quark distribution
$s(x)$ by 
\begin{equation} 
     s(x)=s^{CCFR}(x) +
  \frac{1}{2} \frac{|V_{cd}|^2}{|V_{cs}|^2} \, \delta \bar u(x).  
\end{equation} 
Since $\frac{|V_{cd}|^2}{|V_{cs}|^2}\approx 0.05$, the 
error one makes is roughly two per cent,  
if $\delta \bar u$ is of the same order of magnitude 
as $s(x)$, as the experimental data suggest.  $\delta\bar u(x)$ is 
negative and  hence the true strange quark distribution 
should be smaller than that determined by CCFR neglecting 
charge symmetry violation.  Note that we have neglected all other 
contributions of CSV to the extraction of any other parton 
distributions, and 
we neglect higher order corrections, which could be sizable 
\cite{Barone,Reya}.  

\subsection{Determination of $\mbox{sin}^2(\theta_W )$} 

It might appear that the precision measurement of $\mbox{sin}^2(\theta_W )$ 
from neutrino deep-inelastic scattering, carried out by the CCFR 
Collaboration \cite{SWeinberg}, rules out the possibility of large CSV 
in the parton sea distributions.  Sather \cite{Sather} has previously
pointed out that the measurement of $\mbox{sin}^2(\theta_W )$ is sensitive 
to possible CSV effects in parton distributions. 

If charge symmetry is valid the ratio of the differences of neutrino 
and antineutrino neutral-current and charged-current 
cross sections is given by the Paschos-Wolfenstein 
relation \cite{Paschos} 
\begin{equation} 
 R^- \equiv \frac{\sigma^{\nu N_0}_{NC}
-\sigma^{\bar\nu N_0}_{NC}}
{\sigma^{\nu N_0}_{CC}
-\sigma^{\bar\nu N_0}_{CC}}=\frac{1}{2} 
- \mbox{sin}^2(\theta_W )\, . 
\end{equation} 
The CCFR Collaboration used this relation to extract 
$\mbox{sin}^2(\theta_W )$ and obtained the value  
$\mbox{sin}^2(\theta_W )=0.2255\pm 0.0018(\mbox{stat})\pm 0.0010 
(\mbox{syst})$ \cite{SWeinberg} which is in very good agreement with  
the Standard Model prediction of $0.2230\pm 0.0004$ based on  
measured Z, W and top masses \cite{Data}. The precision of this result  
puts  strong constraints on CSV in parton distributions.  
However, since the  measurement of $\mbox{sin}^2(\theta_W )$   
based on the Paschos-Wolfenstein relation is only sensitive 
to CSV in {\it valence } quark distributions,  
the substantial charge symmetry violation 
in sea-quark distributions found in this analysis does not  
contradict  the precision measurement of $\mbox{sin}^2(\theta_W )$.  

This can be seen as follows. 
The difference between the neutrino and antineutrino charged-current 
cross sections is proportional to the difference between  
$F_2^\nu$ and $F_2^{\bar\nu}$ and to the sum of 
$xF_3^\nu$ and $xF_3^{\bar\nu}$. We see that  
these linear combinations of  
the structure functions are  only sensitive 
to $\delta q(x)-\delta \bar{q}(x)$ i.e. CSV in {\it valence} quark 
distributions (see Eq. \ref{eq2}). 
 
The neutral-current neutrino   cross sections on an iso-scalar target,  
omitting second generation quark contributions, is given by   
\begin{eqnarray} 
  \frac{d\sigma^{\nu N_0}_{NC}}{dxdQ^2} 
 = \frac{G_F^2}{2\pi x} \,\,\, & \{ & 
     a_u \, [u(x)+d(x)-\delta d(x)]x +  
     a_d \,[u(x)+d(x)-\delta u(x)]x+ \nonumber \\  
 & + &     b_u \,[\bar u(x)+\bar d(x)-\delta\bar d(x)]x +  
     b_d \, [\bar u(x)+\bar d(x)-\delta\bar u(x)]x \,\, \} 
\label{nc}
\end{eqnarray} 
Here, we defined $a_f = l_f^2+r_f^2(1-y)^2$ and 
$b_f = l_f^2 (1-y)^2 +r_f^2$ with $f=u,d$ and the couplings 
of the quarks to the neutral-currents are $l_u=1/2-2/3 \,
\mbox{sin}^2(\theta_W)$,  
$r_u=-2/3\, \mbox{sin}^2(\theta_W)$ and 
$l_d=-1/2+1/3\,\mbox{sin}^2(\theta_W)$, 
$r_d=1/3\,  \mbox{sin}^2(\theta_W)$, respectively. 
The antineutrino cross section can be obtained by interchanging 
quarks with antiquarks in Eq. \ref{nc}. 
We immediately see that the difference between neutrino and 
antineutrino neutral-current cross sections is 
only sensitive to CSV in valence quark distributions. 
Thus the large CSV effects in the nucleon sea quark distributions, 
suggested by the CCFR-NMC discrepancy, do 
not influence the measurement of $\mbox{sin}^2(\theta_W)$ 
based on the Paschos-Wolfenstein relation. 

\section{Test of Parton CSV from W Production at Hadron Colliders} 

Clearly, it is important that the charge symmetry violating 
distributions, $\delta\bar d$ and $\delta\bar u$,  
enter with different weights  
in any observable. Otherwise effects due to CSV are not measurable.  
In this connection we also note that 
most of the measured physical observables are proportional 
to the {\it sum} rather than the {\it difference} 
of the charge symmetry violating quark distributions. 
However, $\delta\bar d$ and $\delta\bar u$ are weighted with 
the  charges of the quarks in electro-magnetic interactions,  
such as deep inelastic scattering  
with charged leptons and  Drell-Yan processes. In fact, a comparison 
between charged lepton and neutrino induced structure functions 
was necessary to detect CSV as we have shown in this paper. 
We also discussed the implications of CSV on the   Drell-Yan process.   
In the following, we show that  $W$-boson production in proton 
deuteron collisions can also be used to test the CSV found in this 
paper, if we define a suitable observable. Such measurements could be 
carried out at RHIC and LHC.  Vigdor \cite{Vig97} originally 
suggested that asymmetry in $W$--boson production could be used
as a test of parton charge symmetry.  
  

The cross sections for $pD\rightarrow W^+ X$ 
and $pD\rightarrow W^- X$  are  given by 
\begin{eqnarray} 
  \frac{d\sigma}{dx_F}(pD\rightarrow W^+ X )  
       &\sim & \{u(x_1) [\bar u(x_2)+\bar d(x_2)-\delta\bar u(x_2)] + 
               \bar d(x_1) [u(x_2)+d(x_2)-\delta d(x_2)]\} 
                \mbox{cos}^2\Theta_c \nonumber \\ 
     &+& \{u(x_1)\bar s(x_2) + \bar s(x_1)[u(x_2)+d(x_2)-\delta d(x_2)]\}  
  \mbox{sin}^2\Theta_c\\
 \frac{d\sigma}{dx_F}(pD\rightarrow W^- X )  
       &\sim & \{ d(x_1) [\bar u(x_2)+ \bar d(x_2)-\delta\bar d(x_2)] + 
               \bar u(x_1) [u(x_2)+d(x_2)-\delta u(x_2)]\} 
                \mbox{cos}^2\Theta_c \nonumber\\ 
     &+& \{\bar u(x_1)  s(x_2) +  s(x_1)
   [\bar u(x_2)+\bar d(x_2)-\delta\bar d(x_2)]\} 
        \mbox{sin}^2\Theta_c \,\,. 
\label{wfull}
\end{eqnarray} 
We note that the Cabibbo favored terms in the 
sum  of the $W^+$ and $W^-$ cross sections are invariant 
under the interchange of $x_1$ and $x_2$, if charge symmetry is 
valid. However, 
the Cabibbo suppressed part of the sum contains  terms which are not  
invariant under $x_1\leftrightarrow x_2$, even if charge symmetry is a good 
symmetry. 
Thus, if we define the forward-backward asymmetry as 
\begin{equation} 
   A(x_F) = 
    \frac{(\frac{d\sigma}{dx_F})^{W^+} (x_F) +
     (\frac{d\sigma}{dx_F})^{W^-}(x_F)
    -(\frac{d\sigma}{dx_F})^{W^+}(-x_F)
     -(\frac{d\sigma}{dx_F})^{W^-}(-x_F)}
   {(\frac{d\sigma}{dx_F})^{W^+}(x_F)+ 
    (\frac{d\sigma}{dx_F})^{W^-}(x_F)
   +(\frac{d\sigma}{dx_F})^{W^+}(-x_F)
   +(\frac{d\sigma}{dx_F})^{W^-}(-x_F)} , 
\label{asym}
\end{equation} 
we see that it 
will be  proportional to charge symmetry violating terms and terms 
containing strange quarks.  
Assuming $s(x)=\bar s(x)$ the numerator of Eq. \ref{asym}, 
$\Delta(\frac{d\sigma}{dx_F})(x_F)$
is given by  
\begin{eqnarray} 
\Delta(\frac{d\sigma}{dx_F})(x_F) = \{ 
&  - &[u(x_1)\delta\bar u(x_2)+d(x_1)\delta\bar d(x_2) +  
        \bar u(x_1)\delta u(x_2)+\bar d(x_1)\delta d(x_2)] 
      \, \mbox{cos}^2\Theta_c  \nonumber \\  
  & +& [s(x_1) [d(x_2)+\bar d(x_2) -\delta d(x_2)-\delta\bar d(x_2)]  
\,\mbox{sin}^2\Theta_c \} - (x_1\leftrightarrow x_2) \,.  
\end{eqnarray} 
In the following, we use 
$\delta\bar u \approx -\delta\bar d$, as suggested by our analysis,  
and note that as the charge symmetry violating distribution 
is of the same order of magnitude as the the strange quark distribution,  
 terms proportional to sin$^2\Theta_c$ can be neglected. 
Further, we make the approximations $\delta\bar q(x_2)\approx\delta  q(x_2)$ 
for $x_2\le 0.1$ and $\delta \bar q(x_1)\approx 0$ for large $x_1$. 
We then  obtain 
\begin{eqnarray}  
\Delta(\frac{d\sigma}{dx_F})(x_F) &=&  
   \{ [u(x_1)+\bar u(x_1) -d(x_1)-\bar d(x_1)] \delta\bar d(x_2) 
     \nonumber\\  
   & + & [\delta u(x_1) \bar u(x_2) +\delta d(x_1)\bar d(x_2)] 
        \} \mbox{cos}^2\Theta_c \,\,. 
\label{W} 
\end{eqnarray}
For large $x_F$, 
the forward-backward asymmetry (due to the first term in Eq. \ref{W}) 
is proportional 
to  $\delta\bar d$ times the difference between the  
up and down valence quark distributions. The second term is 
sensitive to CSV in valence quark distributions. 
However, if $\delta d\approx -\delta u$ for valence quarks, 
as suggested by theoretical 
considerations \cite{Lond}, the second term of Eq. \ref{W} is 
approximately $[\bar d(x_2) - \bar u(x_2)] \delta d(x_1)$ 
and is only non-zero if we have FSV. Further, if 
$\delta d(x_1)$ is positive for large $x_1$, as theoretical calculations 
suggest \cite{Lond,Sather}  then 
the second term will contribute positively to the asymmetry, since 
$\bar d -\bar u >0$, so that it would enhance any asymmetry 
expected on the basis of CSV in the sea quark distributions suggested 
by the NMC-CCFR data.  

\begin{figure}
\epsfig{figure=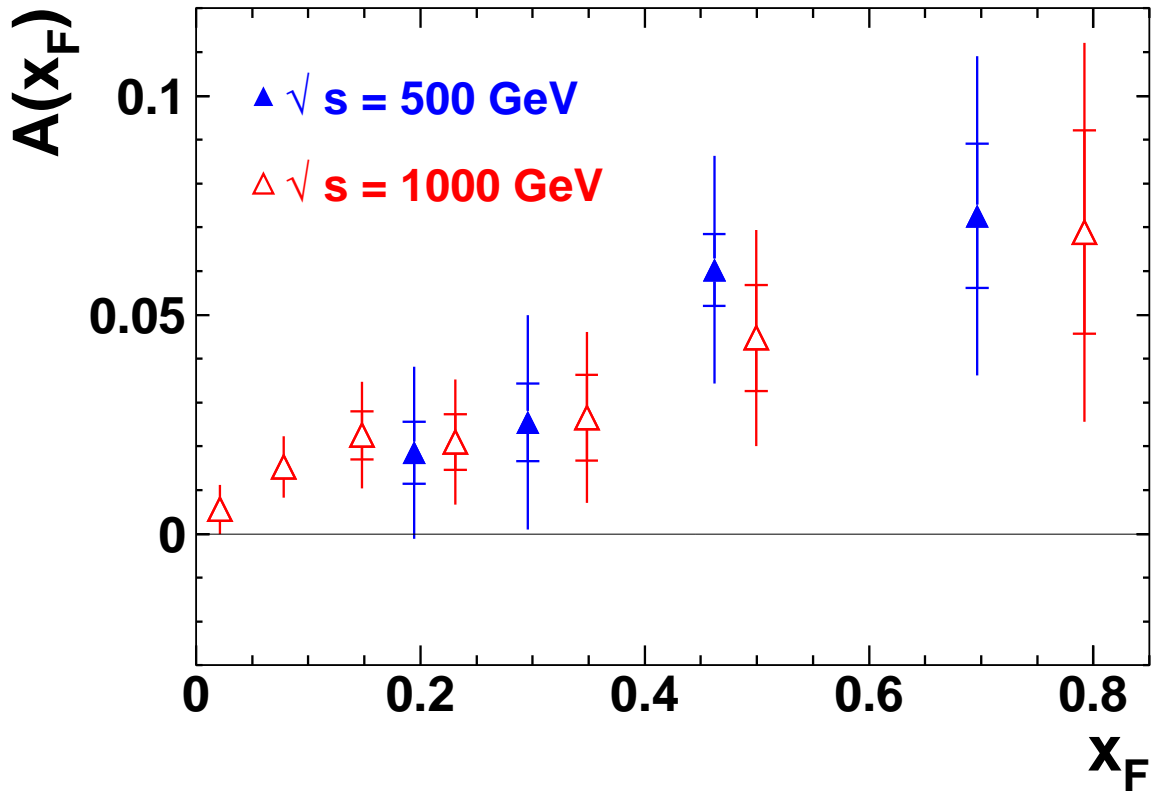,height=14.cm}
\caption{
The forward-backward asymmetry for $W$ production,  
as defined in Eq. 
\protect\ref{asym}. The solid and open triangles 
are calculated for $\protect\sqrt{s}=500$ GeV and  
$\protect\sqrt{s}=1000$ GeV, 
respectively. For $\delta\bar d$,  the values extracted 
from the comparison  of the NMC and CCFR structure function 
are used. The errors are the errors of $\delta\bar d$ and do not 
include the errors of the $W$ experiment. }
\label{fig10}
\end{figure}

We calculated the expected asymmetry $A(x_F)$   
for  $\sqrt{s}=500$ GeV and $\sqrt{s}=1000$ GeV
using the values of $\delta\bar d$ extracted in section II.  
The results are shown in Fig. \ref{fig10}. The error bars represent  
the errors associated with $\delta\bar d$ and do not include 
the errors of the $W$ experiment. 
In the  calculation, we 
retained all terms in Eq. \ref{wfull}.  
The result obtained  by using the approximation in Eq. \ref{W}  
differs only by a few percent from the full calculation. 
We predict considerable asymmetries for large $x_F$.

\section{Conclusions} 

In conclusion, we have examined in detail the discrepancy at 
small $x$ between the CCFR neutrino and NMC muon structure
functions.  Assuming that both the structure functions and
strange quark distributions have been accurately determined 
in this region, we explored the possible reasons for this
discrepancy. First, we re-examined 
the shadowing corrections to neutrino deep inelastic scattering 
and concluded that shadowing cannot account for more than half 
of the difference between the CCFR and NMC structure functions.  
Next, we compared two determinations of the strange quark 
distributions: the ``direct'' method, obtained by measuring
opposite sign dimuon production from neutrino and antineutrino
reactions, and by comparing the CCFR and NMC structure
functions. The strange quark distributions extracted by these
two methods are incompatible with each other, even if we allow 
the antistrange quark distribution to differ 
from the strange distribution in an unconstrained fashion.  

The only way we can make these data compatible is by assuming 
charge symmetry violation in the sea quark distributions.   The
CSV amplitudes necessary to obtain agreement with experiment
are extremely large -- they are of the same order of magnitude
as the strange quark distributions, or roughly 25\% the size of the 
nonstrange sea quark distributions at small $x$.  Such CSV 
contributions are surprisingly large: at least two orders 
of magnitude greater than theoretical predictions of charge
symmetry violation.  We discussed their influence on other observables,  
such as the FSV ratio measured recently in a proton deuteron 
Drell-Yan experiment, on the Gottfried sum rule and on the 
experimental determination of the Weinberg angle 
sin$^2\theta_W$. 
We showed that such large CSV effects could be tested by measuring 
asymmetries in $W$ boson production at hadron colliders such as RHIC 
or LHC.  Such experiments could detect sea quark CSV effects, if they 
were really as large as is suggested by current experiments. 

\vspace*{0.5cm}
\noindent 
{\bf {ACKNOWLEDGMENTS}}
\vspace*{0.5cm}

This work was supported in part by the Australian Research Council.  
One of the authors [JTL] was supported in part by the National
Science Foundation research contract PHY--9722706.  JTL wishes 
to thank the Special Research Centre for the Subatomic Structure
of Matter for its hospitality during the period when this
research was carried out.

\end{document}